\documentclass[journal]{IEEEtran}

\usepackage{amssymb}
\usepackage{lipsum}
\usepackage{amsthm}
\usepackage{subfigure}
\usepackage{gensymb}
\usepackage{textcomp}
\usepackage{multirow}
\usepackage{makecell}
\usepackage{soul,framed} 
\usepackage{url}
\usepackage[table]{xcolor}
\usepackage{cite}
\usepackage{amsmath,amsfonts}
\usepackage{graphicx}
\usepackage{dcolumn}
\usepackage{bm}
\usepackage{pict2e}
\usepackage{epstopdf}
\usepackage{arydshln}
\usepackage{enumitem}
\usepackage[mathscr]{euscript}
\usepackage{algorithmic}
\usepackage{algorithm}
\usepackage{array}
\usepackage{textcomp}
\usepackage{verbatim}
\usepackage{graphicx}
\usepackage{cite}
\usepackage{stfloats}

\definecolor{greencell}{RGB}{200, 255, 200}
\definecolor{yellowcell}{RGB}{255, 254, 174}

\definecolor{bluerevision}{RGB}{0,112,192}

\makeatletter
 \let\old@ps@headings\ps@headings
 \let\old@ps@IEEEtitlepagestyle\ps@IEEEtitlepagestyle
 \def\confheader#1{%
 \def\ps@headings{%
 \old@ps@headings%
 \def\@oddhead{\strut\hfill#1\hfill\strut}%
 \def\@evenhead{\strut\hfill#1\hfill\strut}%
 }%
 \def\ps@IEEEtitlepagestyle{%
 \old@ps@IEEEtitlepagestyle%
 \def\@oddhead{\strut\hfill#1\hfill\strut}%
 \def\@evenhead{\strut\hfill#1\hfill\strut}%
 }%
 \ps@headings%
 }
 \makeatother
\confheader{%
This work has been accepted for publication in \textit{Smart Agricultural Technology}. DOI: 10.1016/j.atech.2025.101260
}

\begin{document}

\title{Multi-Connectivity Solutions for Rural Areas: Integrating Terrestrial 5G and Satellite Networks to Support Innovative IoT Use Cases}

\author{Alejandro Ramírez-Arroyo, Melisa López, Ignacio Rodríguez, Sebastian Bro Damsgaard, Preben Mogensen

\thanks{This work has received funding from the European Union’s Horizon Europe program COMMECT (Under Grant Agreement No: 101060881). The work of Ignacio Rodríguez has been supported by the Spanish Ministry of Science, Innovation, and Universities under Ramon y Cajal Fellowship number RYC-2020-030676-I, funded by MICIU/AEI/10.13039/501100011033 and by the European Social Fund “Investing in your future”. \textit{(Corresponding author: Alejandro Ramírez-Arroyo.)}}

\thanks{Alejandro Ramírez-Arroyo, Melisa López, Sebastian Bro Damsgaard and Preben Mogensen are with the Department of Electronic Systems, Aalborg University (AAU), 9220 Aalborg, Denmark (e-mail: araar@es.aau.dk, mll@es.aau.dk, sbd@es.aau.dk, pm@es.aau.dk).}

\thanks{Ignacio Rodríguez is with the Department of Electrical Engineering, University of Oviedo (UNIOVI), 33203 Gijon, Spain (e-mail: irl@uniovi.es).}}

\markboth{Ramírez-Arroyo \MakeLowercase{\textit{et al.}}: Multi-Connectivity Solutions for Rural Areas: Integrating Terrestrial 5G and Satellite Networks}%
{Ramírez-Arroyo \MakeLowercase{\textit{et al.}}: Multi-Connectivity Solutions for Rural Areas: Integrating Terrestrial 5G and Satellite Networks}

\IEEEpubid{ }

\maketitle

\begin{abstract}
Reliable communications and wide-area coverage are essential factors for innovative applications in rural and remote regions, such as microclimate monitoring, remote operational support, early pest detection, and real-time tracking of livestock transport. However, rural areas often face challenging connectivity conditions due to the lack of terrestrial network (TN) infrastructure, e.g., 5G network technology. While 5G cellular networks are now a reality and promise to improve key performance indicators (KPIs), such as Gbps data rates and latencies in the order of milliseconds, these are typically limited to urban scenarios. To address the rural coverage issue, non-terrestrial networks (NTNs) such as satellite-based solutions, have been introduced to provide coverage in remote regions. Therefore, a multi-connectivity approach can be integrated to simultaneously serve an end-user by merging satellite and cellular links in a joint approach. This study empirically explores the availability of TN (5G) and NTN (satellite) communications in rural areas to develop innovative IoT use cases and applications. It also discusses the possibility of integrating multiple network interfaces through multi-connectivity techniques to improve communication availability and reliability, allowing for the operation of more demanding applications. The results obtained demonstrate that a traditional single-connectivity approach may not be sufficient to provide service to rural environments due to the KPIs requirements given several use cases within these rural areas. TN single-connectivity suffers from constant service outage due to poor infrastructure deployment, while NTN single-connectivity has larger latency and uplink constraint, both limiting the deployment of applications requiring connectivity in rural areas. The multi-connectivity strategy, which jointly integrates 5G and satellite networks, meets the network availability requirements for latency ($< 100$~ms), downlink throughput ($> 30$~Mbps), and uplink throughput ($> 20$~Mbps) KPIs at least 98\%, 99\%, and 95\% of the time, respectively. Therefore, multi-connectivity solutions enable several use cases, such as precision agriculture, livestock monitoring, and forest management, which are unfeasible with traditional single-connectivity approaches.
\end{abstract}


\section{Introduction}
\label{introduction}

Rural areas present several unique use cases which require highly reliable communication networks to provide adequate service. For instance, there are multiple use cases, e.g., crop and livestock monitoring~\cite{crop_monitoring,livestock_monitoring}, weather data management~\cite{weather_management}, or precision agriculture~\cite{precision_agriculture}, where network connectivity is essential for the proper provisioning of the resources required for each task. However, despite the factual deployment of 5G cellular networks that promise the enhancement of several key performance indicators (KPIs) capable of meeting end-user's needs, such as several Gbps data rates and latencies in the order of few milliseconds \cite{survey_5G, survey_6G}, rural areas remain underserved. This fact can be attributed to the sparse population density and the high cost of infrastructure deployments, where despite efforts to extend these terrestrial networks to non-densely populated areas~\cite{survey_rural_area_1}, there exists a coverage penetration issue in rural areas since the deployment of cellular networks in these scenarios is not cost efficient~\cite{survey_rural_area_2}. In order to address this issue, solutions based on non-terrestrial networks (NTNs) have been developed in recent years, more specifically relying on satellite coverage \cite{6G_NTN, satellite_launch}. Thus, satellite networks are able to provide coverage in remote areas where cellular networks are not deployed \cite{satellite_networks}. Currently, there are already commercial options that offer these services~\cite{starlink}.

In addition, the concept of multi-connectivity between cellular networks has recently emerged, where two independent networks (typically with independent infrastructure) are combined to provide service to the same end-user. This concept offers the advantage of using each of the operators separately in a joint network approach~\cite{MC_cellular_1, MC_cellular_2}. This allows to leverage the network coverage of multiple operators simultaneously, maximizing throughput and minimizing network delays \cite{MC_latency}. Thus, TN-TN (5G-5G) multi-connectivity can be established by taking advantage of simultaneous use of the coverage of two mobile network operators. One step beyond, two technologies with different architectures can be integrated, leading to the development of \mbox{TN-NTN}~(\mbox{5G-Sat.}) multi-connectivity, where both networks are used simultaneously to increase the reliability \cite{survey_TN_NTN, integration_TN_NTN_spectrum}. For instance, the terrestrial link can be operated when available, while the satellite link is maintained as a backup in case the terrestrial link is not operational \cite{integration_TN_NTN}. Note that the joint integration of multiple infrastructures may bring not only the aforementioned advantages, but also a series of challenges such as Doppler shift, protocol and architecture heterogeneity, or inter-network mobility and handover management, which should be carefully considered during the integration design.

Based on the state of the art, theoretical studies on multi-connectivity have primarily focused on cellular networks, with a few exploring satellite networks, and are often based on analytical or simulation models. Therefore, there are few studies focused on solutions for remote areas, as well as a lack of empirical validations of this type of solution for rural use cases. For instance, Kibria \textit{et al.}~\cite{SOA_MC_simulation_Hetnets} develop a stochastic model for dual connectivity evaluation performance in heterogeneous cellular networks. Majamaa \textit{et al.}~\cite{SOA_packet_duplication} examine several packet duplication techniques to maximize resource allocation efficiency without compromising network performance. Wolf \textit{et al.}~\cite{SOA_MC_reliability} and Weedage \textit{et al.}~\cite{SOA_MC_capacity} focus on the analysis of network KPIs such as capacity or outage probability to evaluate the benefits of multi-connectivity in cellular networks. Additionally, research is also carried out on the effect of multi-connectivity in specific use cases and scenarios, such as the case of dynamic blocking in mmWave systems \cite{SOA_MC_blockage}, or its reliability in vehicular environments \cite{SOA_MC_V2X}. In terms of multi-connectivity between terrestrial and non-terrestrial networks, Li \textit{et al.}~\cite{review_2}~present several deployment architectures for satellite-terrestrial integrated networks, as well as system design challenges, solutions, and applications. In addition, Shang \textit{et al.}~\cite{review_1} propose a theoretical framework for integrating both networks with MIMO systems, determining downlink coverage probability and average achievable data rate when combining both technologies. From an experimental perspective, fewer contributions explore multi-connectivity solutions for rural or infrastructure-limited areas. There have been efforts in the development of prototypes of integrated cellular and satellite systems \cite{SOA_5GALLSTAR}. Similarly, proofs of concept based on the validation of cellular multi-connectivity for Internet of Farming \cite{SOA_IoF}, or for public protection and disaster relief \cite{SOA_disaster} have been proposed. Despite these efforts, empirical validations of multi-connectivity in rural environments remain scarce, particularly those involving heterogeneous technologies such as 5G and satellite.

In this work, an experimental characterization of the benefits of multi-connectivity in rural environments for both TN-TN (5G-5G) multi-connectivity and TN-NTN (5G-Sat.) multi-connectivity is carried out. For this purpose, an analysis of the physical layer of the environment based on parameters such as outage probability or signal level is performed. In parallel, the analysis of the network layer reveals the performance of KPIs such as throughput or latency that have a direct impact on the end-user's experienced quality of service. Thus, the analysis of both layers allows an in-depth evaluation of multi-connectivity as a tool to improve network reliability. Our previous works assessed the data rate for single-connectivity cases and latency for TN-NTN multi-connectivity for static user equipment (UEs) \cite{AAU_1}, and latency for TN-NTN multi-connectivity for dynamic UEs \cite{AAU_2}. This work performs an assessment on cellular/satellite coverage and service availability, as well as a joint comprehensive evaluation of multi-connectivity between cellular-cellular and cellular-satellite networks in both the physical layer (signal strength and outage probability) and the network layer (throughput and latency) under dynamic conditions in rural scenarios. Therefore, the work novelty lies in jointly exploring cellular-cellular and cellular-satellite multi-connectivity through experimental and empirical evaluation in rural environments, addressing a range of KPIs aligned with realistic IoT and agricultural use cases. A key contribution of this study is the extensive measurement campaign carried out under real-world conditions, which constitutes one of the few comprehensive empirical campaigns reported in this field. Therefore, this study yields quantitative values in terms of the performance of a communications network operating in multi-connectivity mode versus a traditional network connectivity operating in single-connectivity mode. This contribution focuses on improving connectivity in areas where commercial solutions exist but are limited by poor infrastructure, thus providing a robust and resilient multi-connectivity approach based on existing commercial connectivity services. The satellite-cellular and cellular-cellular solutions assessment offer a series of guidelines on integrating these technologies to enhance network performance in agricultural use cases. The main contributions of this work are:

\begin{itemize}

\item Identification of a set of use cases and types of connectivity that require network access in rural environments to develop innovative IoT use cases and applications.

\item Identification of a representative rural area of 20~$\textrm{km}^2$ and development of an extensive measurement campaign. The campaign includes (i) passive measurements, which analyze the physical radio layer and the propagation scenario, thus identifying the main radio propagation
challenges encountered when establishing communications in rural environments, as well as (ii) representative live-data transmissions KPIs such as latency or throughput, which directly impact the quality of service for end-users.

\item Development of a dedicated measurement setup, which allows data to be acquired from two 4G/5G cellular mobile networks and a Starlink satellite link simultaneously, enabling the study of TN-TN (5G-5G) and \mbox{TN-NTN (5G-Sat.)} multi-connectivity cases.

\item Analysis of empirical measurement results and identification of the main network availability implications and connectivity challenges to operate rural IoT use cases given single-connectivity and multi-connectivity solutions.

\item Mapping the measured network performance to rural application needs, identifying which use cases can be enabled and how they can operate under different connectivity conditions based on latency and data rate availability.

\end{itemize}

The combination of previous contributions presents an experimental framework to guide the deployment of connectivity solutions. Therefore, the results are expected to demonstrate the benefits of multi-connectivity communications, thereby facilitating the development of integrated wireless systems incorporating this technology. Likewise, this study focuses on applying this technology to rural use cases, as these are among the most impacted by limited network performance in purely terrestrial deployments due to a lack of adequate infrastructure. Note that the solution presented in this work is not aimed at creating new coverage spots based on new infrastructure, but rather at increasing the robustness and resilience of connectivity based on existing networks. Demonstrating a stable network connection and meeting minimum KPI requirements enables the development of use cases that enhance rural areas by increasing productivity and promoting development in rural contexts. Contrary to studies based exclusively on theoretical approaches, this work presents a quantitative and experimental evaluation carried out in a representative rural environment in northern Europe that includes both satellite and terrestrial networks with limited coverage. The selected area is characterized by sparse population, limited land infrastructure, intermittent coverage, and traffic patterns typical of rural contexts. These characteristics are representative of many rural regions across Europe. As such, the results can be extrapolated to other rural regions with comparable connectivity infrastructures. Consequently, the study offers generalizable conclusions on network performance improvements enabled by multi-connectivity configurations, serving as a practical reference for future deployments.

This work is organized as follows. Section~II introduces a set of use cases related to rural areas that could be impacted by the proposed experimental framework. Section~III presents and describes the equipment employed for the measurement campaign, the environment under analysis, and the multi-connectivity tool that implements this live approach. Section~IV evaluates the different rural IoT use cases identified, and Section~V provides a summary on their feasibility under the different network operation options explored. Finally, the main conclusions of the work are drawn in Section~VI.

\section{Use cases and performance requirements in rural scenarios}

This Section categorizes the types of communications that can occur between interconnected devices in rural areas, and it identifies a set of use cases in the context of the need for connectivity in rural areas and agricultural work in the countryside. This identification has been carried out based on the work developed under European Union initiatives, such as IoF 2020 and COMMECT projects \cite{IoF, COMMECT}. From a general point of view, three main types of communication have been identified based on the devices involved:

\begin{enumerate}

\item \textbf{Machine-to-Data Center Communications (M2DC)}: It consists of the exchange of data between integrated devices in rural areas with data centers. This exchange usually occurs before or after the operational period of the rural tasks, both for pre-processing and post-processing data. Communication is bidirectional both downlink and uplink, and examples of the data shared include high-resolution images and videos, maps or software updates of the devices themselves.

\item \textbf{Machine-to-Smart Device Communications (M2SC)}: This type of communication is related to the exchange of information between the devices deployed in the rural area and the smart devices of the agricultural workers. Thus, workers can carry out actions and real-time monitoring of agricultural tasks. Some data shared in this category include images, monitoring data or instructions from the worker in real time.

\item \textbf{Machine-to-Machine Communications (M2M)}: These communications are carried out between the devices deployed in the rural area. They are related to the exchange of data between devices for the correct sensor network operation. They refer to small amounts of data linked to the monitoring of the rural environment, such as soil conditions, weather data, machinery monitoring, or energy consumption.

\end{enumerate}

In the context of this work, the term \textit{machine} refers to any automated or sensor-equipped device deployed in rural areas, such as agricultural drones, weather stations, irrigation systems, or autonomous tractors. Meanwhile, a \textit{smart device} denotes personal or portable electronic devices used by agricultural workers, including smartphones or tablets that facilitate real-time interaction with the deployed machines.

Beyond the types of communications, up to nine specific use cases are identified where TN-NTN (5G-Sat.) network integration could be beneficial to improve deployment and network reliability. These nine use cases are considered representative of the diversity of connectivity conditions required in rural areas and how the use of multi-connectivity solutions can enhance the KPIs involved. For instance, note that the use cases later addressed present scenarios such as monitoring of agricultural areas (e.g., vineyards or olive groves), pest control, monitoring of emergency conditions in forests, or livestock transport. These use cases represent a wide range of examples that illustrate the need for connectivity under different circumstances in rural and remote regions. In more detail, these are as follows:

\begin{itemize}

\item \textbf{Microclimate and crop monitoring in vineyards (UC1)}: It aims to optimize vineyard protection by monitoring the field with leaf wetness and temperature sensors. This allows winegrowers to make informed decisions about the timing and dosage of treatments, improving accuracy and reducing unnecessary fungicide applications.

\item \textbf{Digital twin for digitalized vineyard management (UC2)}: This use case focuses on creating a digital twin of vineyards by integrating data from various sensors, remote sensing, and IoT systems. The digital twin enables precise monitoring and management of vineyard operations, including disease detection, irrigation scheduling, and fertilization, by providing real-time spatial accurate data.

\item \textbf{Remote operational support in forests (UC3)}: It involves providing remote guidance and control for forest operations through technologies like VR, AR, and digital cameras. It aims to enhance decision-making, reduce errors in logging and thinning activities, and improve machine maintenance by enabling remote experts to provide real-time support.

\item \textbf{Complex situation awareness service in the forests (UC4)}: This use case focuses on providing emergency personnel with real-time digital information during critical situations like accidents, fires, or floods in forested areas. By employing celullar networks, with satellite networks as backup, it aims to enhance situational awareness and enable quick action responses. The technology supports the efficient management of emergencies, improving operator safety and decision-making in rapidly changing conditions.

\item \textbf{Monitoring of livestock transport along routes (UC5)}: This use case addresses the need for continuous monitoring of livestock transport units, enabling real-time reporting of vehicle locations and sensor data to several operation centers. It aims to enhance logistics efficiency by providing truck drivers with internet access for route optimization, including updates on traffic conditions, weather forecasts, and disease risk areas.

\item \textbf{License plate recognition (UC6)}: Automated license plate recognition enables online verification of certificates for vehicles entering farms. This process involves scanning the license plate, checking the associated certificates in real-time against regulatory databases, and automatically granting access. It requires reliable connectivity to function properly in rural scenarios.

\item \textbf{Monitoring livestock loading/unloading processes (UC7)}: Automated systems for counting, weighing, and illness detection in livestock during loading and unloading aim to reduce manual labor and errors by utilizing computer vision technology. Cameras positioned on trailer ramps and in delivery rooms capture images for real-time processing. Effective deployment of this technology requires robust connectivity to support the live streaming and transmission of data, ensuring accurate and efficient monitoring in rural areas.

\item \textbf{Early disease and pest detection in olive crops (UC8)}: Early warning systems based on weather data and pathogen-environment models help olive farmers optimize disease and pest control. By monitoring conditions such as temperature, humidity, and soil moisture, these systems provide alerts for interventions, like fungicide spraying, to prevent the spread of diseases. The implementation of these technologies in rural areas requires reliable D2D connectivity to ensure continuous data transmission from weather stations.

\item \textbf{Monitoring of pest insect traps (UC9)}: The use of digital traps equipped with cameras and wireless connectivity enables remote monitoring of several insect populations in olive groves. These traps, which capture images and transmit data continuously, allow for timely interventions without the need for frequent manual inspections. Effective deployment of this technology relies on robust wireless network connectivity to ensure continuous data transmission.

\end{itemize}


\renewcommand{\arraystretch}{1.5}

\begin{table}[!t]
\centering
\caption{Network connectivity requirements identified for the set of connectivity types and use cases identified in rural areas. The parameters displayed have been derived as part of the studies performed in IoF2020 and COMMECT European projects.}
\label{tab:use_cases}
\resizebox{1\columnwidth}{!}{%
\begin{tabular}{|cccc}
\hline\hline
\multicolumn{1}{|c||}{} & \multicolumn{1}{c|}{Availability (Reliability)} & \multicolumn{1}{c|}{Latency} & \multicolumn{1}{c|}{Throughput (DL/UL)} \\ \hline\hline
\multicolumn{4}{|c|}{\textbf{Main types of communication - general application}}                                                                                           \\ \hline\hline
\multicolumn{1}{|c||}{M2DC}  & \multicolumn{1}{c|}{99\% ($10^{-2}$)}   & \multicolumn{1}{c|}{100 ms}   & \multicolumn{1}{c|}{50/50 Mbps} \\ \hline
\multicolumn{1}{|c||}{M2SD} & \multicolumn{1}{c|}{99\% ($10^{-2}$)}   & \multicolumn{1}{c|}{100 ms}   & \multicolumn{1}{c|}{10/10 Mbps} \\ \hline
\multicolumn{1}{|c||}{M2M}   & \multicolumn{1}{c|}{99\% ($10^{-2}$)}   & \multicolumn{1}{c|}{100 ms}   & \multicolumn{1}{c|}{5/5 Mbps}   \\ \hline\hline
\multicolumn{4}{|c|}{\textbf{Identified rural IoT use cases - with integrated TN-NTN potential}}                                                                                                       \\ \hline\hline
\multicolumn{1}{|c||}{UC1 \textit{(Precision agriculture)}}      & \multicolumn{1}{c|}{99\% ($10^{-2}$)}   & \multicolumn{1}{c|}{400 ms}  & \multicolumn{1}{c|}{5/5 Mbps}   \\ \hline
\multicolumn{1}{|c||}{UC2 \textit{(Digital Twin)}}      & \multicolumn{1}{c|}{99\% ($10^{-2}$)}   & \multicolumn{1}{c|}{400 ms}  & \multicolumn{1}{c|}{1/5 Mbps}   \\ \hline
\multicolumn{1}{|c||}{UC3 \textit{(Remote Assistance)}}      & \multicolumn{1}{c|}{99.9\% ($10^{-3}$)} & \multicolumn{1}{c|}{100 ms}  & \multicolumn{1}{c|}{1/20 Mbps}  \\ \hline
\multicolumn{1}{|c||}{UC4 \textit{(Emergency Response)}}      & \multicolumn{1}{c|}{99.9\% ($10^{-3}$)} & \multicolumn{1}{c|}{100 ms}  & \multicolumn{1}{c|}{1/20 Mbps}  \\ \hline
\multicolumn{1}{|c||}{UC5 \textit{(Transport Logistics)}}      & \multicolumn{1}{c|}{99.9\% ($10^{-3}$)} & \multicolumn{1}{c|}{400 ms}  & \multicolumn{1}{c|}{14/1 Mbps}  \\ \hline
\multicolumn{1}{|c||}{UC6 \textit{(Automated Access)}}      & \multicolumn{1}{c|}{99\% ($10^{-2}$)}   & \multicolumn{1}{c|}{1000 ms} & \multicolumn{1}{c|}{1/5 Mbps}   \\ \hline
\multicolumn{1}{|c||}{UC7 \textit{(Computer Vision)}}      & \multicolumn{1}{c|}{99\% ($10^{-2}$)}   & \multicolumn{1}{c|}{1000 ms} & \multicolumn{1}{c|}{5/10 Mbps}  \\ \hline
\multicolumn{1}{|c||}{UC8 \textit{(Predictive Analysis)}}      & \multicolumn{1}{c|}{90\% ($10^{-1}$)}   & \multicolumn{1}{c|}{400 ms}  & \multicolumn{1}{c|}{5/5 Mbps}   \\ \hline
\multicolumn{1}{|c||}{UC9 \textit{(Remote Sensing)}}      & \multicolumn{1}{c|}{90\% ($10^{-1}$)}   & \multicolumn{1}{c|}{400 ms}  & \multicolumn{1}{c|}{1/5 Mbps}   \\ \hline\hline

\end{tabular}%
}
\end{table}


The requirements in terms of connectivity for communications and the previously presented use cases are shown in Table~\ref{tab:use_cases}. These are a combination of three factors required to ensure the quality of service in each case: (i)~latency, conceived as the network round-trip time (RTT) to establish communications between two devices, (ii)~network throughput, both in uplink and downlink, since depending on the use case these values might be asymmetric, and (iii)~network availability, understood as the time in which the network is up and it is possible to meet the previous latency and throughput requirements.

In general, the examples described throughout this Section require mobile devices in rural areas where they can provide connectivity services, whether, for example, in the process of harvesting vineyards and olive trees, transporting livestock, or forest management. This mobility, where the quality of connectivity conditions can change significantly in short periods, is taken into account in planning the measurement campaign and the empirical analysis introduced in Sections~III and IV. Additionally, the previous use cases will serve as a baseline in Section~V to quantify and validate the advantages of integrating various connectivity technologies to reduce the digital divide in rural areas. Note that the connectivity criteria in the nine use cases may align with the three connectivity types outlined. They are presented separately because, while the connectivity types provide an overview of the data volume required based on the hardware involved, the use cases emphasize the practical utility of deploying a connectivity solution.

\section{Measurement scenario, network setup and multi-connectivity tool, and measurement procedures}

This Section describes the rural scenario under study, the network measurement setup (including multi-connectivity tool), as well as the tests considered in the measurement campaign.

\subsection{Measurement scenario}

\begin{figure}[!t]
	\centering
	\includegraphics[width= 1\columnwidth]{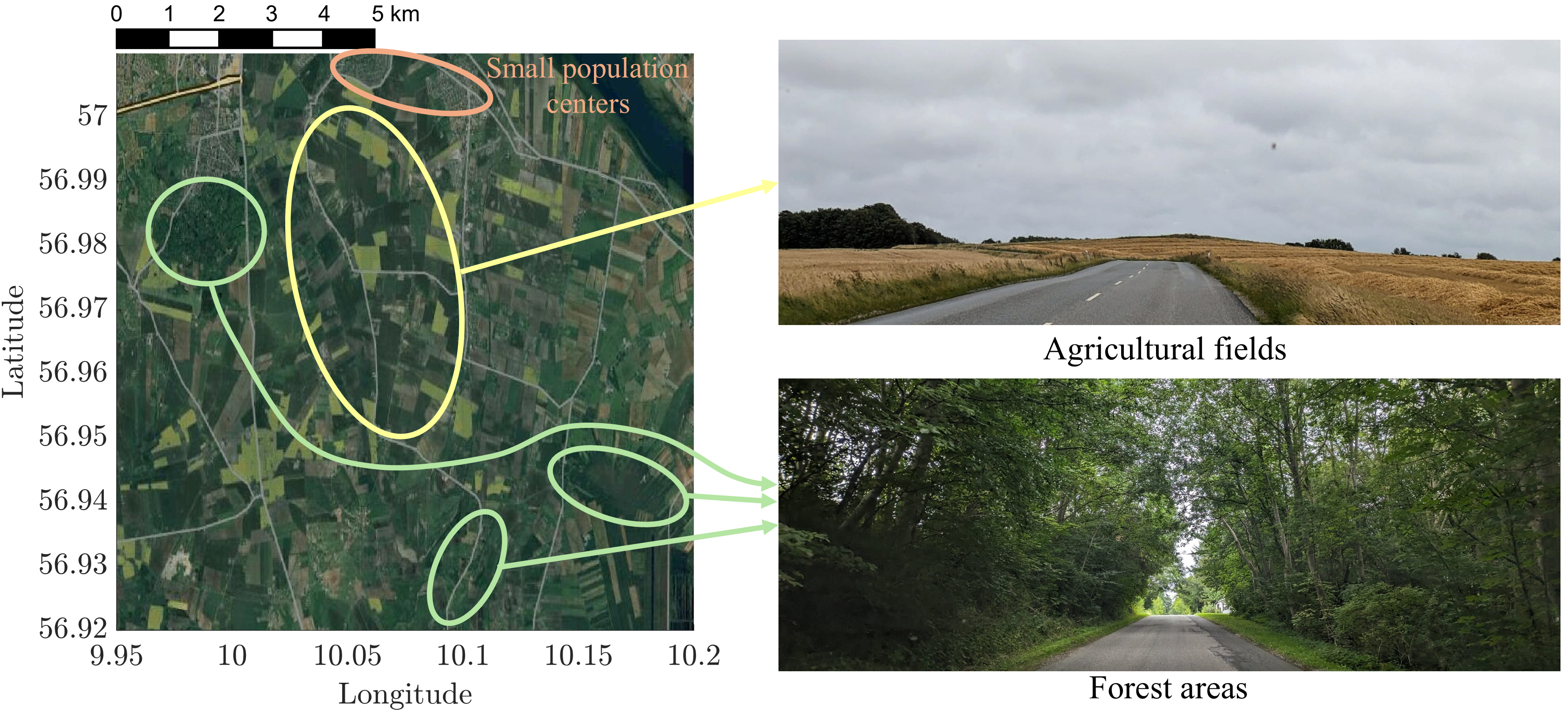}
    \caption{Aerial overview of the selected areas for the measurement campaign, which includes small population centers, agricultural fields, and forest areas. Most representative environments have been recorded during the measurement campaign.} 
    \label{landscape}
\end{figure}

\begin{figure}[!b]
	\centering
	\includegraphics[width= 1\columnwidth]{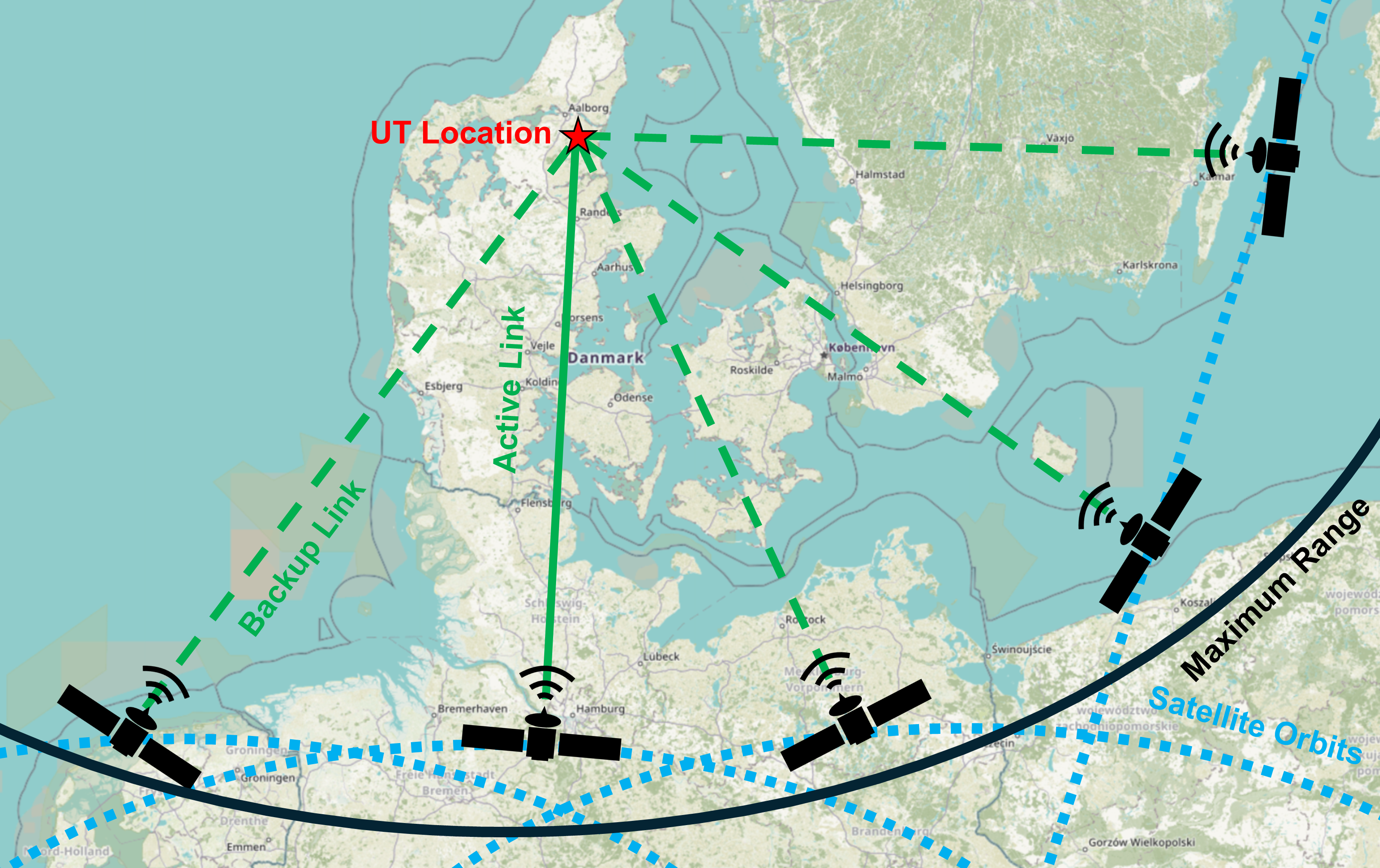}
    \caption{Illustration of the typical satellite availability in Aalborg (Denmark). UT typically reaches between 3 and 7 satellites. These follow orbits with $53\degree$ inclination crossing northern Germany and polar orbits across Sweden.} 
\label{fig_sim}
\end{figure}

The study is performed in a rural area located between 10~km and 20~km south-east of the city of Aalborg (Denmark). This region, which is driven for 30~km over a maximum area of 20~$\textrm{km}^2$, has no fluctuations in ground height of more than 50 meters and it is predominantly composed of farms, cultivated areas, and small population centers. Hence, the typical landscape along the driven route consists of wide cultivated plains, as well as wooded areas, illustrated in Fig.~\ref{landscape}. The measurement area under study is selected as a representative sparsely populated rural region, characterized by (i) limited land infrastructure, (ii) intermittent coverage, (iii) and traffic patterns typical of rural areas. These conditions are commonly observed in many rural scenarios of Europe, making the results applicable beyond the specific location~\cite{SOA_IoF, COMMECT}. Thus, the previous factors are relevant for evaluating the resilience and performance of multi-connectivity solutions in real rural scenarios. Both Line-of-Sight (LoS) and Obstructed Line-of-Sight (OLoS) conditions can be expected between the UEs and the base station/satellite for TN/NTN networks, respectively. To emulate the behavior of some of the previous presented use cases in Section~II, such as the movement of agricultural vehicles performing several tasks in the field, driving is conducted at low speeds regime along the route, specifically at 20 km/h.

Regarding the cellular network availability, it is deployed through several base stations (BSs) working in the 700~MHz, 800 MHz, 900 MHz, 1800 MHz and 2100 MHz frequency bands for 4G LTE technology, and 700 MHz and 1800 MHz for 5G NR technology. According to the Danish Agency for Digital Government~\cite{SDFI}, the highest BS density is deployed in the northern area, corresponding to the outskirts of the city of Aalborg, where small towns can be found. Meanwhile, the southernmost area has a lower BS density due to the predominant presence of farm fields (see Section~IV). Concerning the satellite network, Starlink has approximately 4,500 satellites deployed in LEO orbit by the time of measuring, and is currently the satellite network with the highest broadband service worldwide. Given the location of the measurement campaign to be performed and the orbits established by the satellites, Starlink provides coverage to the city of Aalborg, mostly from satellites navigating in northern Germany and a few satellites navigating Sweden in polar orbits. Most orbits providing coverage to the region under study are based on Walker Delta constellations with latitudes of around 53º~\cite{degrees_orbits}. Given the number of satellites and the elevation angle required to establish a ground-satellite connection, the minimum number of satellites available is typically between 3 and 7, which ensures continuous connectivity if Line-of-Sight is guaranteed, and redundant connectivity even if one of the visible satellites is unavailable. Fig. \ref{fig_sim} shows an illustration of the satellites available at a specific time.

\subsection{Measurement setup}

\begin{figure}[t]
	\centering
	\subfigure[]{\includegraphics[width=1\columnwidth]{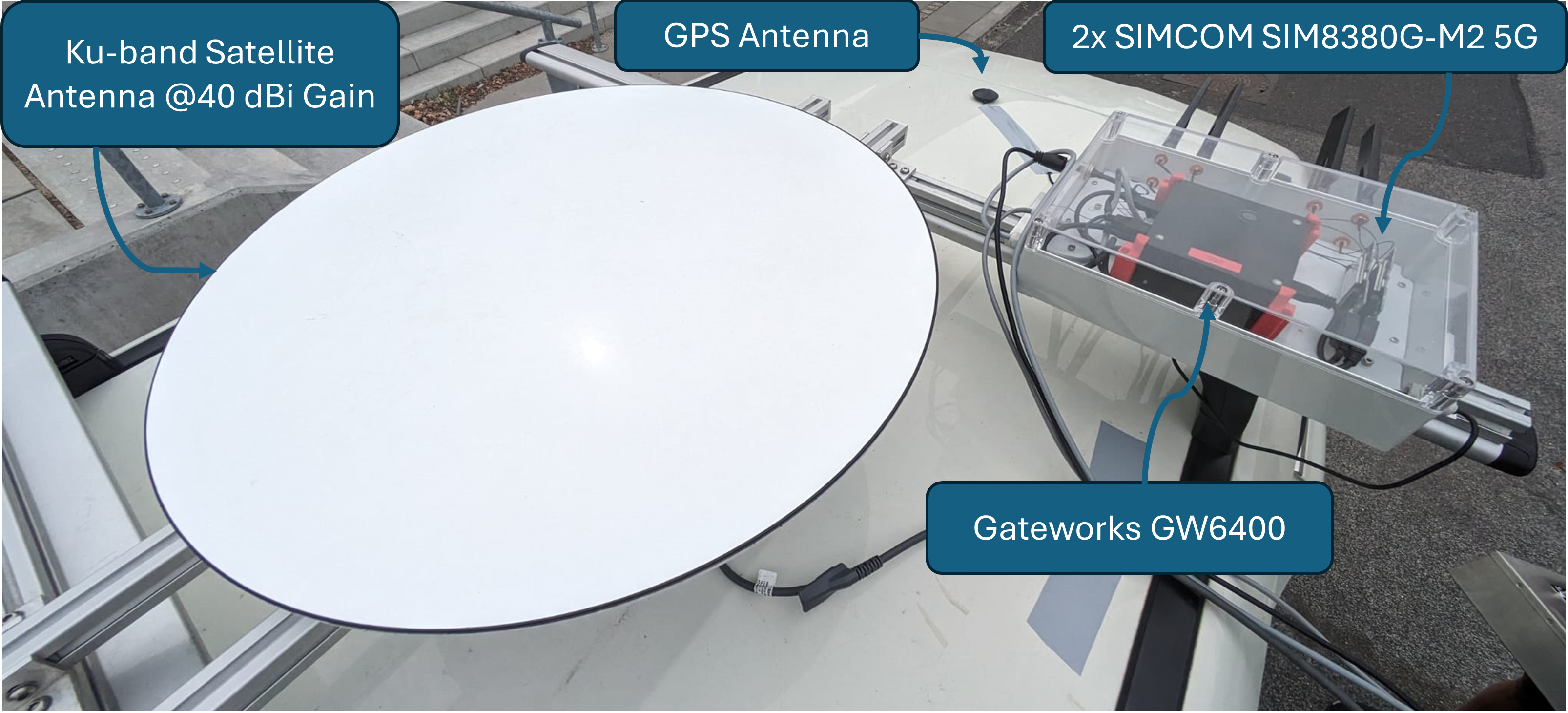}
	} 
\subfigure[]{\includegraphics[width= 1\columnwidth]{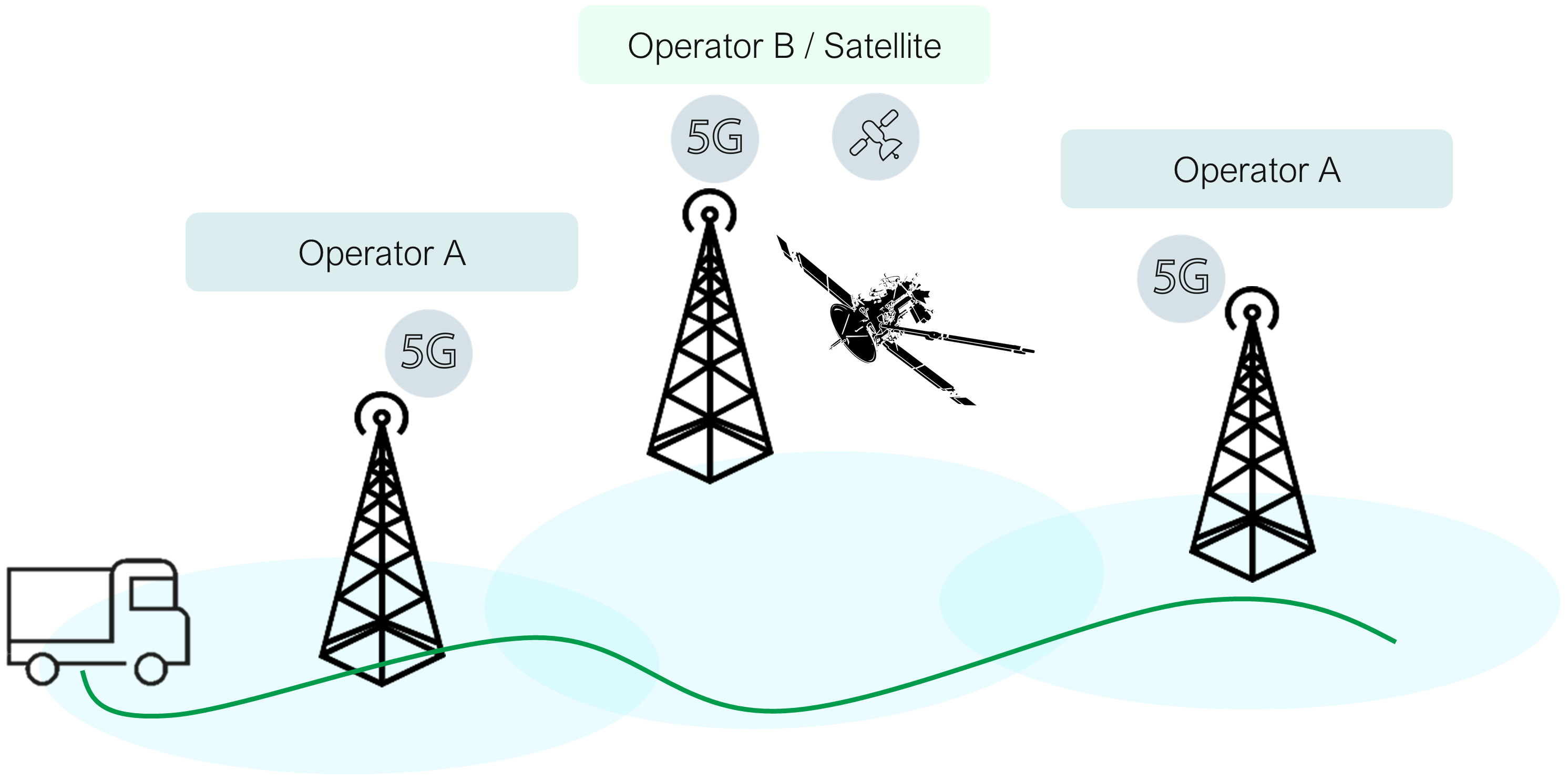}
	}
	\caption{(a) Measurement equipment deployed for the experimental characterization of multi-connectivity networks and (b) conceptual use case.} 
	\label{setup}
\end{figure}

To assess the multi-connectivity capabilities, a dedicated setup has been assembled. Specifically, the processing system consists of a single-board computer GW6400 \cite{GW6400}. Attached to this gateway are two multi-band SIM8380G-M2 modules with support for 5G NR/LTE-FDD/LTE-TDD/HSPA+ technologies \cite{SIMCOM}. The modules contain SIM cards from two different Danish operators with the densest 4G/5G deployments in the country. In addition, a Starlink Gen-1 Ku-Band Satellite antenna is also attached to provide connectivity to the Starlink satellite network deployed in the low Earth orbit (LEO) \cite{Starlink_UT}. Finally, this setup is assembled on the roof of a car for the measurement campaign. Additionally, a GPS antenna is included to keep track of the location of the measurements. The complete assembly is illustrated in Fig.~\ref{setup}(a). Thus, the setup emulates a vehicular entity with connectivity to three different wireless interfaces, two of them cellular networks, hereafter called Operator A and Operator B, and a satellite network provided by the Starlink constellation. Fig. \ref{setup}(b) shows a conceptual use case where our vehicle is primarily covered by the coverage of operator A given a specific route to be covered between two locations. However, due to the sparse deployments of cellular networks in rural environments, there are coverage holes where operator A is not able to provide service. Therefore, a connection is established with a second cellular network (operator B) or a non-terrestrial network (satellite) that acts as a backup link to prevent service interruption. Note that this conceptual framework is intended to generally represent the majority of the previously presented use cases, in which mobility, coverage holes, or both might be found in the scenario given several use cases.

\begin{figure*}[!t]
	\centering
	\includegraphics[width= 0.8\textwidth]{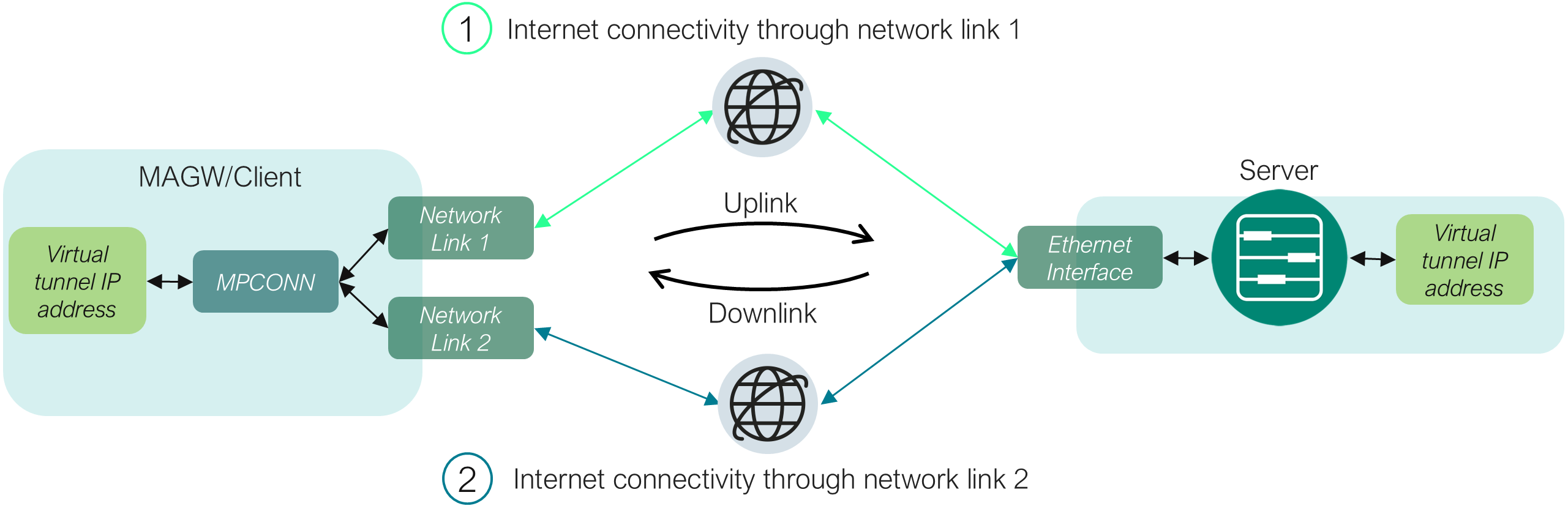}
    \caption{Client-server schematic to establish multi-connectivity through \textit{mpconn} tool.} 
	\label{fig4}
\end{figure*}

In order to test and demonstrate the benefits of multi-connectivity between mobile and satellite technologies, Aalborg University has developed a tool for packet duplication \cite{mpconn}. This tool duplicates IP packets on the network layer, tunneling each packet through a series of pre-established network interfaces. Because of the tool tunnels and performs packet duplication at the IP layer, this process is completely transparent to network architectures, which are unaware that packets on one network are being transmitted through other interfaces and vice versa. This is essential as it avoids problems related to synchronization between network architectures and any other physical layer and access layer aspects. Fig. \ref{fig4} illustrates a simplified scheme of the tool. From the client side, represented by the GW6400 single-board computer, a connection is established with a server. Any IP packet generated at the client is forwarded through this tool, which duplicates each packet and transmits it over two interfaces: a cellular interface and a satellite interface, or two different cellular interfaces. A virtual tunnel is created between the client and server via the available interfaces. In the case of the cellular network the traffic is routed via BSs deployed in the field, which are directly connected to the Internet. On the other hand, the satellite traffic is routed via the satellite constellation, which forwards the communication flow through terrestrial gateways. On the server side, also running the multi-connectivity tool, the first packet reaching the server side will be captured, as well as the interface it was received from. This server setup is deployed at Aalborg University using a 1 Gbps Ethernet interface. The duplication procedure is completely analogous in both uplink and downlink, and the impact of the tool on latency and jitter is practically negligible ($< 1$~ms), as it only involves additional encapsulation and decapsulation at the IP layer~\cite{mpconn_impact}. This arrangement allows for the evaluation of network performance under single-connectivity, multi-connectivity cellular-cellular, and multi-connectivity cellular-satellite scenarios. In this work, this tool operates in full duplication mode, meaning that all packets are transmitted simultaneously over the two available interfaces. This approach maximizes multi-connectivity performance, as packets are delivered through the fastest interface, regardless of network conditions. This configuration is intended to showcase the upper performance bound achievable with multi-connectivity. However, as it will be discussed in Section~V, future implementations may involve smart packet duplication strategies based on live network KPIs.

\begin{figure*}[!t]
	\centering
	\includegraphics[width= 1\textwidth]{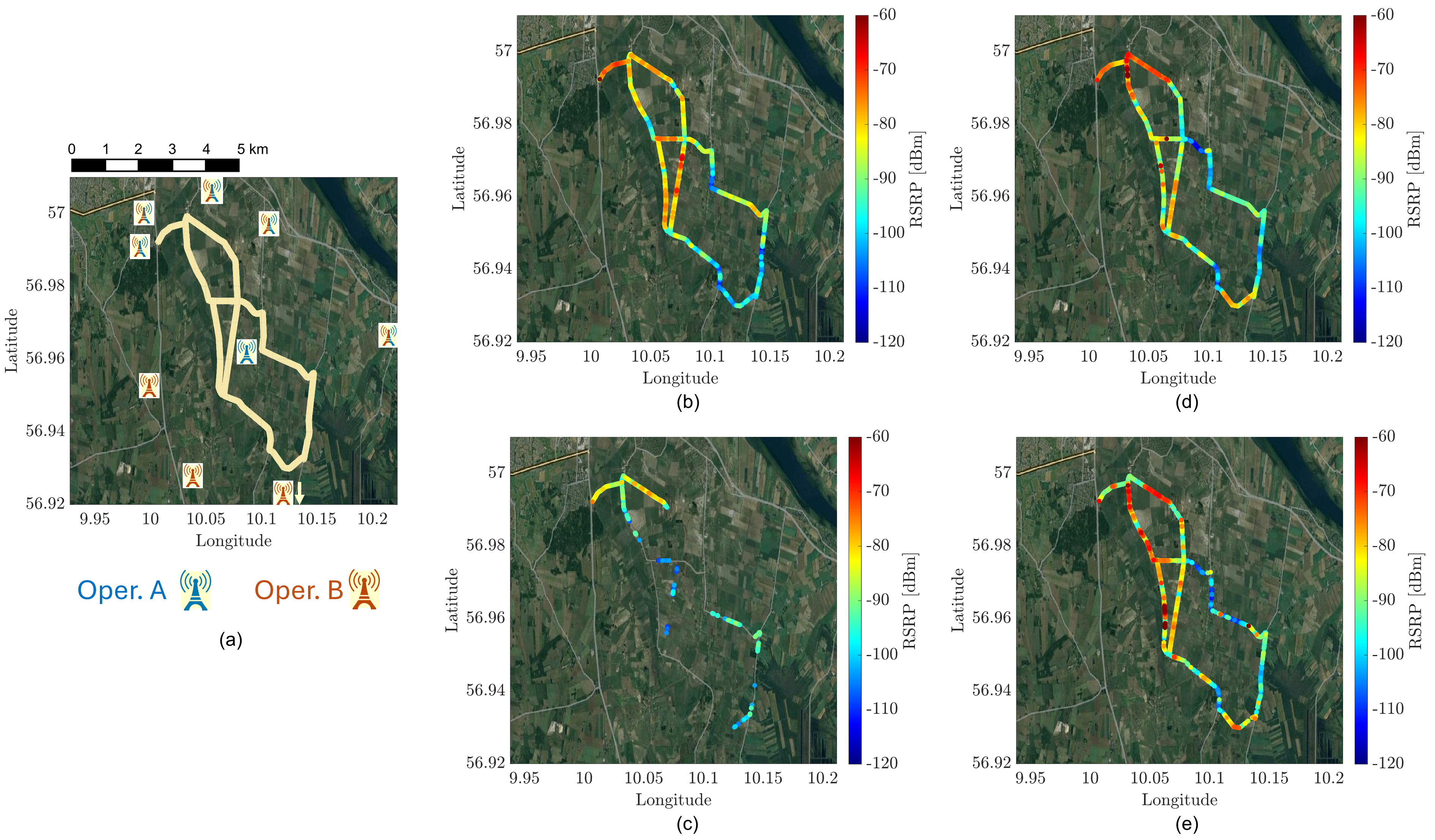}
	\caption{(a) Driven route through the rural area under tests. Cellular BSs for both operators are illustrated. RSRP levels are measured for (b) 4G LTE - Operator A, (c) 5G NR - Operator A, (d) 4G LTE - Operator B, and (e) 5G NR - Operator B.} 
	\label{fig_route}
\end{figure*}

\subsection{Measurement procedures}

Throughout Section~IV, network performance will be analyzed based on four KPIs: (i)~coverage availability, (ii)~latency, (iii)~downlink throughput, and (iv)~uplink throughput. Likewise, three possible connectivity configurations are analyzed: (a)~single-connectivity, (b)~cellular-cellular multi-connectivity, and (c)~cellular-satellite multi-connectivity. The route established and the exact location of the terrestrial network BSs of operators A and B in the vicinity of the route, which takes into account the area under study in Fig. 1, is illustrated in Fig.~\ref{fig_route}(a). Note that in some cases, both terrestrial operators share the same radio mast, while in other cases, the mast is exclusively served by a single operator. Since each KPI and technology must be analyzed separately, given the setup shown in Section~III.B, several route iterations shown in Fig.~\ref{fig_route}(a) are performed. In each iteration, the route is driven for a total of 2~hours, thus allowing for a statistical analysis of the data obtained.

To measure the cellular 4G/5G coverage availability from the physical layer perspective, the signal strength of the cellular network is set as the Reference Signal Received Power (RSRP) which is the power level averaged over 127 Synchronization Signal Block (SSB) subcarriers in 5G \cite{Alex_FR2}, or averaged over the Cell Specific Reference Signal (CRS) signal in 4G. Regarding the satellite network, frequency bands and orbits are public through the U.S. Federal Communications Commission \cite{Starlink_license}, due to the commercial nature of this network, multiple data on the technology used remains confidential. However, several efforts are currently underway to understand the signal structure employed \cite{Starlink_structure}. The transmission between the User Terminal (UT) provided by Starlink and the satellite occurs in the Ku-band between 10.7~GHz and~12.7~GHz. This communication is redirected to one of the multiple deployed Starlink terrestrial gateways. Note that in the non-terrestrial network, it is not possible to measure satellite coverage through passive measurements as it is performed with the terrestrial network, which is mainly due to Starlink's proprietary protocols. However, active measurements (latency and throughput) are carried out in connected mode, which are detailed in Sections~IV.B and IV.C.

For the latency analysis, we set up a test based on the networking tool \textit{ping} where this command is run every 100~ms with 64B packet size. The total time measured is the round-trip time from when the client sends the request until it receives a response from the server. To compare the advantages of multi-connectivity against single-connectivity, we first evaluate single connectivity for both cellular links and the satellite link. Additionally, multi-connectivity is assessed between both cellular networks (MC - Cellular-based) and between the cellular network of operator A and the satellite network (MC - Cellular-Satellite). As described in Section~III.B, our multi-connectivity tool duplicates and launches the \textit{ping} command simultaneously through the selected interfaces, considering the successful transmission to be the one that reaches the server side with the lowest delay. Note that it is established that latencies exceeding 2~s are considered service outages, and the packets are discarded since this threshold significantly deviates from the expected values according to the standards for cellular and LEO satellite technologies.

For the throughput assessment, we use the \textit{iperf3} tool where we establish UDP connections with our server given a target symmetrical bandwidth of 100 Mbps (100/100 Mbps DL/UL). This bandwidth value is chosen as it far exceeds the typical values required in rural environments presented in Section~II, so that the network can be analyzed when pushing it towards high capacity values. Due to the behavior of the traffic sent through \textit{iperf3}, whose output is the average throughput of all packets sent through the network in 1~s intervals, independent traffic requests are made through each interface and the one with the maximum throughput is considered optimal in the multi-connectivity scenario. As in the latency study, the tests were conducted in the environment illustrated in Fig.~\ref{fig_route}(a), with driving routes of up to 2 hours, leading to 7200 samples per interface averaged over~1~s. This analysis includes both downlink and uplink throughput for single-connectivity and multi-connectivity strategies.

\section{Experimental results}

This Section details the main results of the performed measurement campaign in terms of cellular coverage availability, and latency/throughput for cellular, satellite, and multi-connectivity network solutions.

\subsection{Cellular 4G/5G coverage availability in the targeted rural area}

\begin{figure}[!t]
	\centering
	\includegraphics[width= 1\columnwidth]{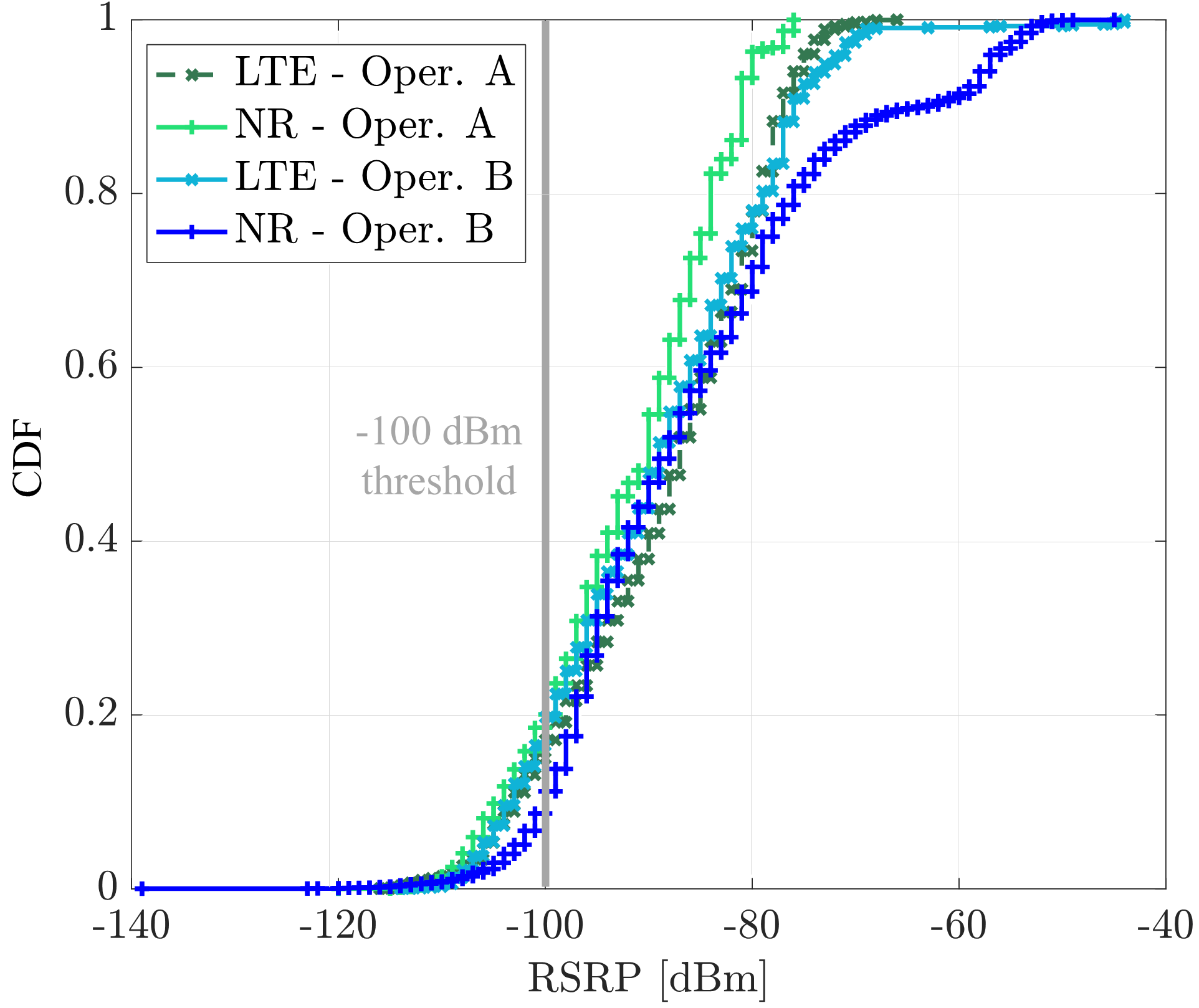}
	\caption{Cumulative distribution function for RSRP values given 4G LTE and 5G NR deployments for operators A and B.} 
	\label{fig5}
\end{figure}

From a physical layer perspective, the signal strength analysis provides an overview of the coverage conditions where the campaign is conducted. Thus, RSRP is measured for the two technologies analyzed given operators~A~and~B. Figs. \ref{fig_route}(b) and \ref{fig_route}(c) show the signal level for operator A in 4G LTE and 5G NR technologies respectively, while Figs.~\ref{fig_route}(d) and \ref{fig_route}(e) present these RSRP levels for operator B. These figures show that the areas with the highest RSRP levels are concentrated in the north, with a higher concentration of BSs due to the population centers. However, as we move towards the southern part of the route, the signal levels decrease due to the few base stations that cover the region. This fact is especially noticeable for operator A's 5G network, which is only visible in the northern part of the map and intermittently in some locations along the route. There is also a visual relation between signal levels and the location of each operator's base stations. For instance, in the southernmost driven region, signal levels are notably higher for operator B due to the deployment of BS in this area exclusively by this operator. Fig. \ref{fig5} illustrates the CDF for the RSRP given the cellular technologies and operators available. The median RSRP values are very similar, ranging from $-86.9$~dBm to $-90.0$~dBm. However, noticeable differences are observed in the upper part of the CDF where the highest signal strengths are found. Considering the top 10\% signal strength, operator B stands out with RSRP levels above $-63.3$~dBm and $-75.9$~dBm for 5G NR and 4G LTE, respectively. In comparison, these values drop to $-80.9$~dBm and $-77.0$~dBm for operator A, respectively. In addition, Table~II shows some of the statistics related to this evaluation of the physical layer in the cellular network. In this table, it can be observed the low availability of the 5G network for operator A of only 35.7\%, as previously illustrated in Fig.~\ref{fig_route}(c). Assuming that a user can be served by any of the technologies given one of the operators, operator A has a null probability of being out of coverage due to the 100\% availability of the 4G network. However, operator B has some holes of coverage due to the fact that the 4G network is not always available, coinciding with temporary instants where the 5G network is not available either, having a 0.2\% probability of not having either network available along the route. Note that the standard deviation $\sigma$ of the signal strength is higher for operator B, as illustrated in Table~II. This fact is because operator B has the highest signal levels which explains the high variability.

\renewcommand{\arraystretch}{1.5}

\begin{table}[!t]
\centering

 \caption {Network coverage statistics in terms of RSRP for the two operators and cellular technologies}  \label{table1} 
\resizebox{0.8\columnwidth}{!}{

\begin{tabular}{|cc||c|c|}
\hline \hline
                                              &                            & Operator A        & Operator B         \\ \hline \hline
\multicolumn{1}{|c|}{\multirow{3}{*}{5G NR}}  & Availability               & 35.7\%            & 74.5\%             \\ \cline{2-4} 
\multicolumn{1}{|c|}{}                        & RSRP mean                  & -91.7 dBm         & -85.7 dBm          \\ \cline{2-4} 
\multicolumn{1}{|c|}{}                        & RSRP $\sigma$              & 8.6 dB            & 13.5 dB            \\ \hline \hline
\multicolumn{1}{|c|}{\multirow{3}{*}{4G LTE}} & Availability               & 100\%               & 99.8\%             \\ \cline{2-4} 
\multicolumn{1}{|c|}{}                        & RSRP mean                  & -88.4 dBm         & -88.9 dBm          \\ \cline{2-4} 
\multicolumn{1}{|c|}{}                        & RSRP $\sigma$              & 9.8 dB            & 10.8 dB            \\ \hline \hline
\multicolumn{2}{|c||}{Out of coverage}                                       & 0\%               & 0.2\%              \\ \hline \hline
\end{tabular}
}
\end{table}

Despite the high availability of cellular networks, it should be noted that low signal strength levels can have a significant impact on network performance due to the low SNR regime in which communications occur. For a critical RSRP value of $-100$ dBm \cite{Melisa_Access}, in the best case (5G NR - Oper. B) there is still a 9.8\% chance that samples will be below $-100$ dBm, while this percentage increases for other cases to 16.0\% (4G LTE - Oper. A), 18.3\% (4G LTE - Oper. B), and 19.1\% (5G NR - Oper. A). Note that these values only include those samples where the signal was available given the sensitivity of the equipment. For higher equipment sensitivity, the previous percentages would have increased significantly.

In summary, the previous results concerning the physical layer show a non-negligible probability that network conditions may not be optimal for establishing an Internet connection in the rural scenario studied. Thus, a single cellular operator may not be sufficient to offer continuous and seamless service, either due to: (i) coverage gaps due to insufficient 5G network deployment by operator A, (ii) areas without coverage for 4G and 5G networks by operator B, or (iii) low RSRP values in some parts of the route. For these reasons, this scenario is considered suitable for the analysis of multi-connectivity techniques, either through the combined use of cellular operators or the use of a satellite link as a backup.

\subsection{Latency analysis of cellular, satellite, and multi-connectivity network solutions in connected-mode}

\begin{figure}[!t]
	\centering
	\includegraphics[width= 1\columnwidth]{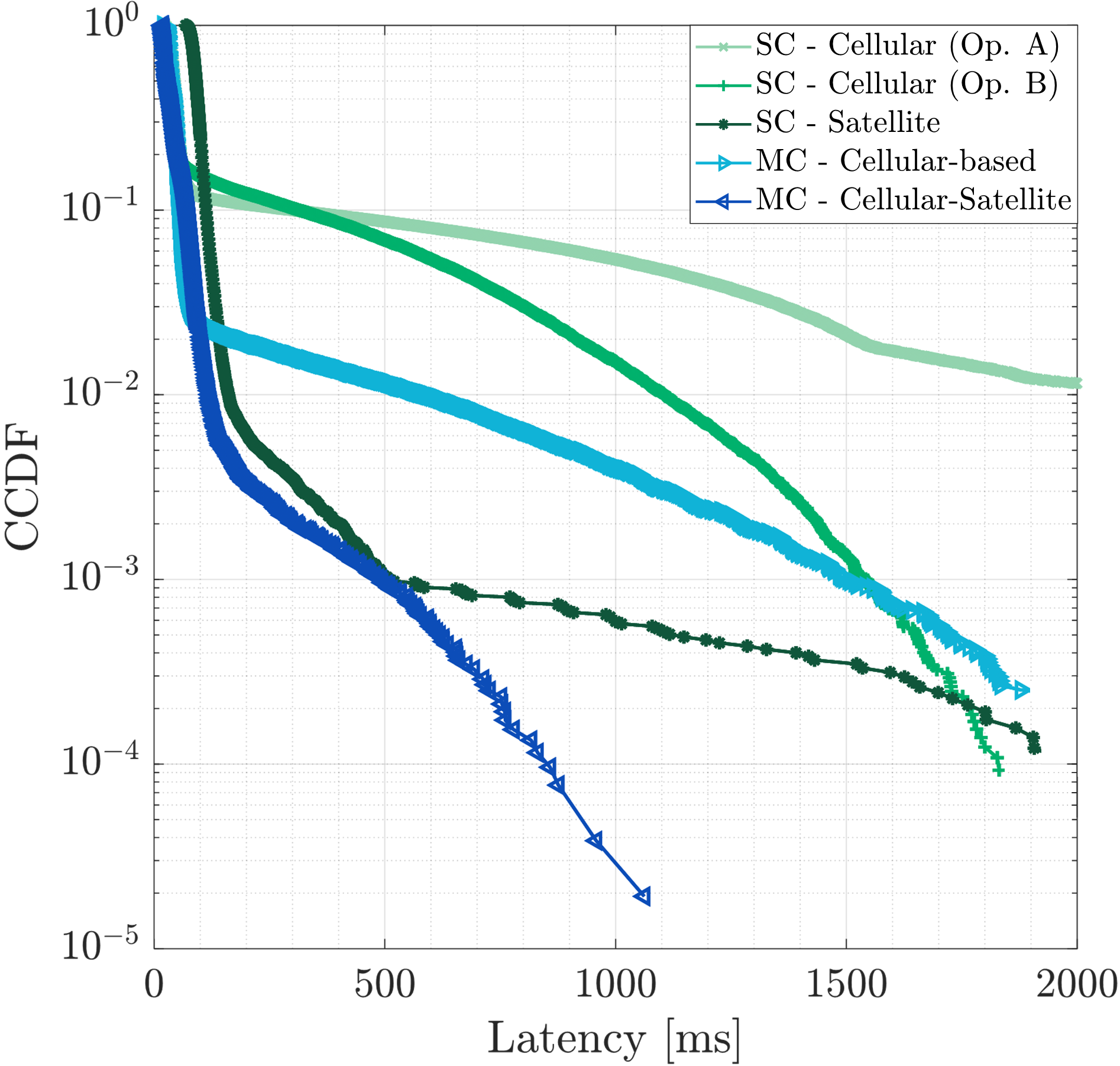}
	\caption{Complementary cumulative distributed function for latency values given single-connectivity and multi-connectivity approaches in rural scenarios. Latency values larger than 2000~ms are considered a service outage.} 
	\label{fig6}
\end{figure}

In this Section, we analyze the latency according to the different connectivity options available (see Section III.C). Table~III shows the main statistics of the study conducted and Fig.~\ref{fig6} displays the complementary cumulative distribution function (CCDF) for each deployment.

\renewcommand{\arraystretch}{1.5}
\setlength{\tabcolsep}{4pt}

\begin{table*}[!b]
\centering
\caption{Latency statistics for single-connectivity and multi-connectivity approaches in rural scenarios}
\label{table2}
\resizebox{\textwidth}{!}{%
\begin{tabular}{|c||c|c|c|c|c||c||c|c|c|}
\hline
\hline

                         & Min     & Median  & Mean      & Max     & Std. dev & Service outage & 99\%             & 99.9\%           & 99.99\%          \\ \hline \hline
Operator A               & 24.7 ms & 44.8 ms & 130.2 ms  & 1999 ms & 295.3 ms & 1.1\%          & \textgreater 2 s & \textgreater 2 s & \textgreater 2 s \\ 
Operator B               & 20.1 ms & 28.9 ms & 105.2 ms  & 1831 ms & 219.2 ms & 0.008\%        & 1103 ms          & 1541 ms          & \textgreater 2 s \\ \
Satellite                & 68.8 ms & 90.8 ms & 95.7 ms   & 1971 ms & 41.1 ms  & 0.009\%        & 161 ms           & 508 ms           & \textgreater 2 s \\ \hline
Cellular-based MC        & 19.6 ms & 31.2 ms & 48.1 ms   & 1982 ms & 104.5 ms & 0.02\%         & 582 ms           & 1503 ms          & \textgreater 2 s \\ 
Cellular-Satellite MC    & 17.5 ms & 31.7 ms & 41.3 ms   & 1062 ms & 32.3 ms  & 0\%            & 118 ms           & 489 ms           & 851 ms           \\ \hline \hline
\end{tabular}
}
\end{table*}

Concerning the single-connectivity analysis, operator A resulted in 1.1\% of samples where latency is above 2~s, potentially due to several coverage gaps in the rural environment. For the operator B and the satellite case, these service outages are around 0.1\% ($10^{-3}$ in the tails of the CCDF). The median values for operators A and B are 44.8~ms and 28.9~ms, respectively, while it is 90.8~ms for Starlink technology. Note that there is a significant variation between the mean and median in the cellular cases, primarily due to the large standard deviation noticed in cellular technology (see Table~III). Visually, this can be observed in the sharp increase in the tails of the CCDF starting at approximately $10^{-1}$, i.e., the worst 10\% of cases. These values can be associated with signal values in the physical layer below $-100$~dBm, whose percentage $(\sim10\%-20\%)$ is equivalent to that observed in latency samples with notable degradation in the tails. Thus, operators A and B may be unsuitable given the KPIs requirements previously presented in Table~\ref{tab:use_cases} due to the observed tail in the CCDF. Concerning the satellite link, the median and mean values are similar, where this median increases by about 45-60 ms compared to the cellular operators. This effect can be attributed to the longer propagation delays between the nodes involved in the communication, and also the signal processing time at each network node, i.e., user terminal antenna, LEO satellite and ground station. In return, the link is much more robust in terms of jitter, as the CCDF decreases more rapidly compared to both cellular cases.

For the multi-connectivity approaches, it can be observed that the use of cellular-based MC maintains the median latency around the best case obtained in single connectivity. The main improvement is found in the $>100$ ms region due to the robustness introduced by the availability of both interfaces. For the same reason, this robustness in the link reduces the jitter to 104.5 ms, which is below the results of the single connectivity cases for each operator. Similarly, the service outage is on the same order of magnitude as the best single connectivity case, i.e., $10^{-3}$. Note that since the single-connectivity and multi-connectivity tests were conducted at different time instants, there is statistical variability in the channel that causes fluctuations in the tails of each test conducted. This implies that in the very high latency regime, this statistical variability may result in better values for single connectivity than for multi-connectivity, as observed in Fig.~\ref{fig6}. However, a trend towards convergence between both tails can be seen. For the cellular-satellite MC case, it is noticeable the lowest average and jitter latency value among all the studied cases. This can be attributed to the cellular link generally being used as the main link under regular conditions. However, the satellite link remains on standby as a backup, substituting the cellular link when the tails spike, as observed in the single-connectivity case. This case is the only one analyzed that offers seamless connectivity since it was observed that at least one of the two networks was always available along the route. Finally, the CCDF for cellular-satellite MC shows the best latency values, where for 99\%, 99.9\%, and 99.99\% of cases, the latency is below 118 ms, 489 ms, and 851 ms, respectively, which is lower than the previous cases as shown in Table~III.

In summary, the results suggest that multi-connectivity between cellular and satellite networks offers the best results in terms of latency, with a network availability of 100\%. Analyzing the records of the multi-connectivity tool, operator~A was the best network 44.4\% of the time, while operator~B was the best 55.6\% of the time in the case of cellular-based MC. These results suggest that there is no clear dominance of either network and that resources are equally shared. This leads to the intermediate region of the CCDF, where multi-connectivity performs better than either link alone. However, in the high latency regime, we observe a convergence towards the best single-connectivity case. In the cellular-satellite MC case, the cellular link is employed 86.2\% of the time, while the satellite link is used only 13.8\% of the time. This indicates that the cellular interface provides the primary link. Nevertheless, when high latencies occur, the backup satellite network satisfactorily covers the link. This combination has proven to be the best of the proposed cases, demonstrating through an empirical approach the feasibility of multi-connectivity approaches to decrease latency in areas with low cellular network penetration.

\subsection{Throughput analysis of cellular, satellite, and multi-connectivity network solutions in connected-mode}

\begin{figure}[!b]
	\centering
	\includegraphics[width= 1\columnwidth]{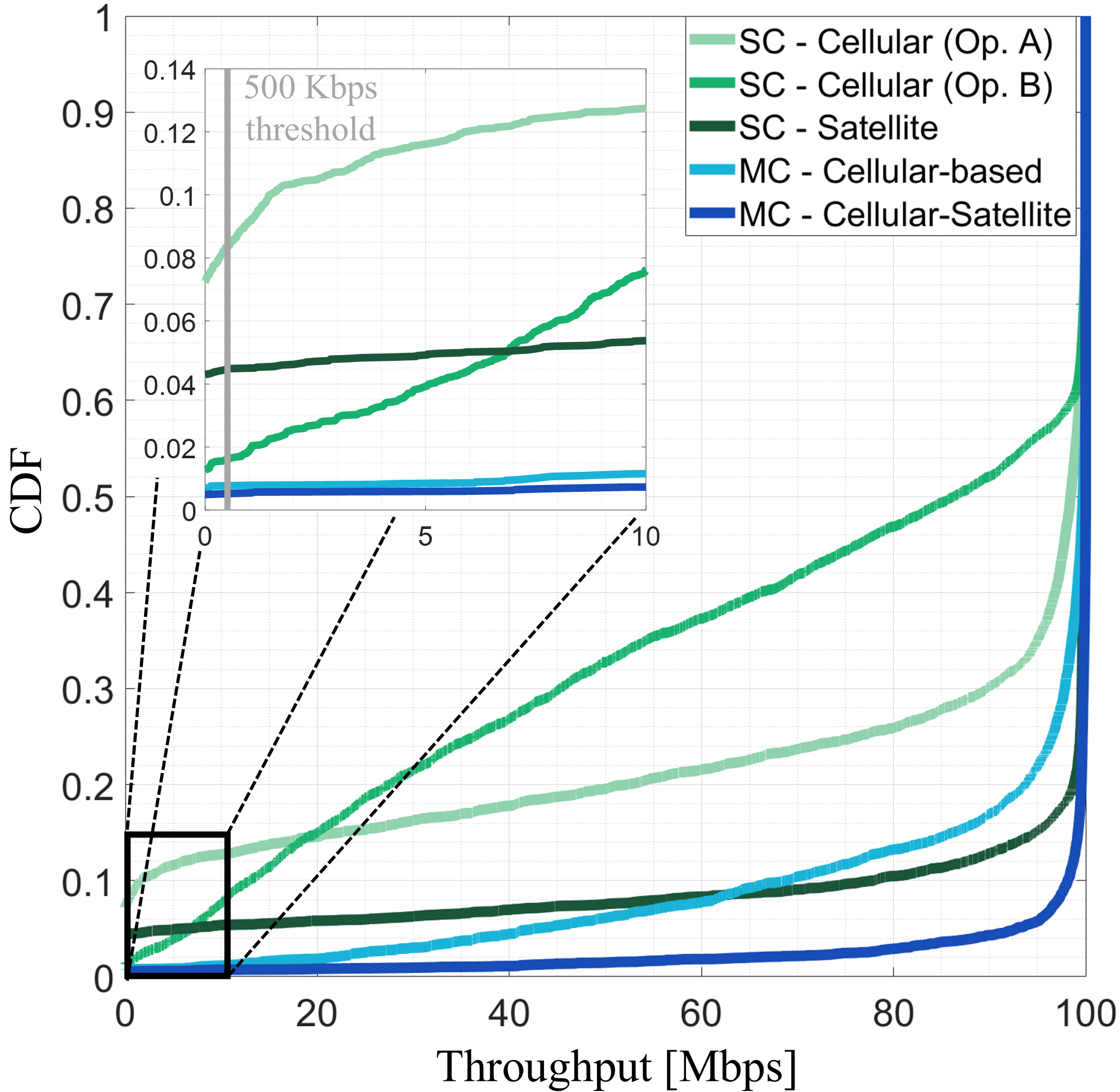}
	\caption{Cumulative distributed function for downlink throughput values given single-connectivity and multi-connectivity approaches in rural scenarios. Downlink throughput values below 500~Kbps are considered a service outage.} 
	\label{figDL}
\end{figure}

\begin{figure}[!b]
	\centering
	\includegraphics[width= 1\columnwidth]{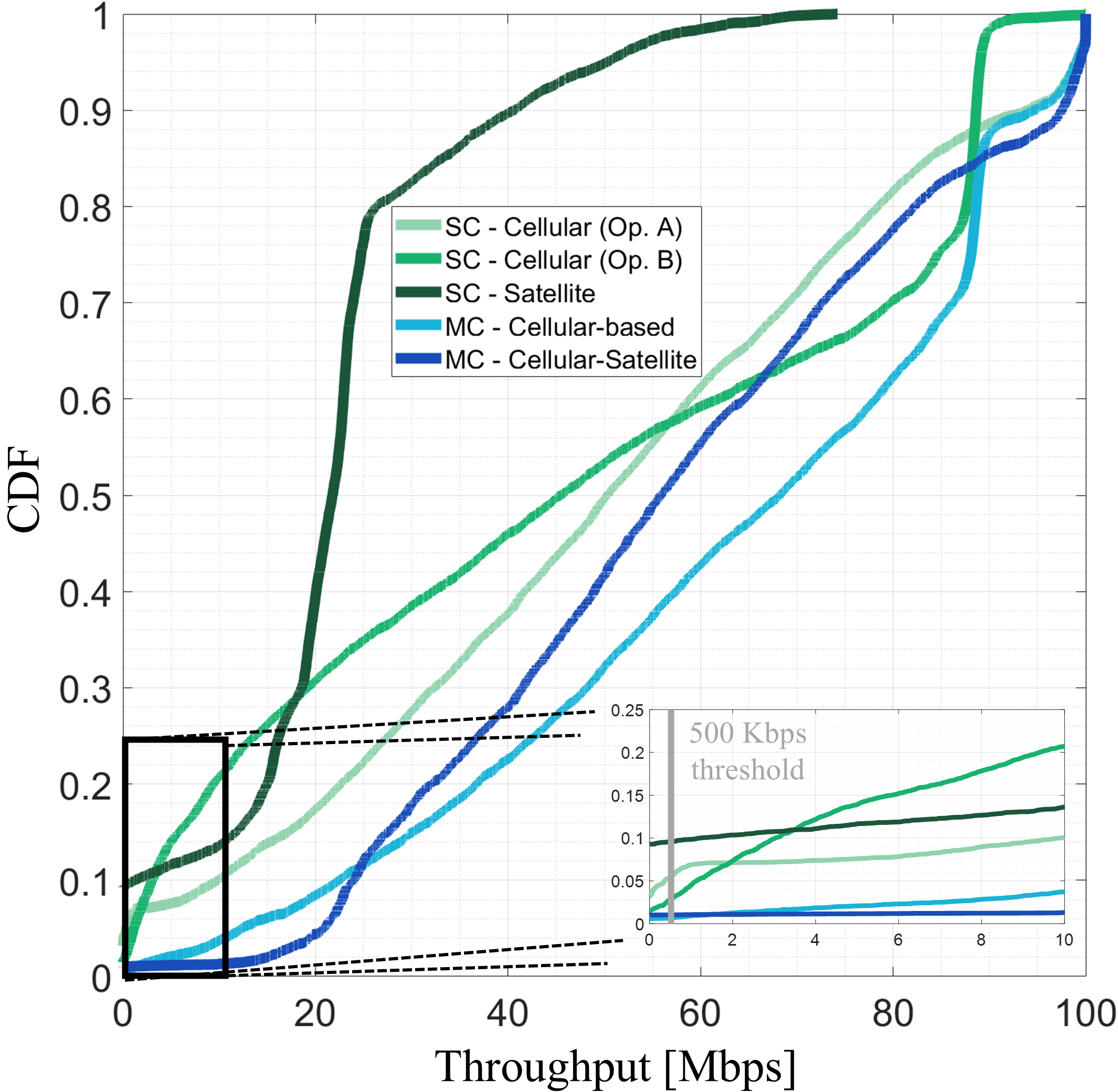}
	\caption{Cumulative distributed function for uplink throughput values given single-connectivity and multi-connectivity approaches in rural scenarios. Uplink throughput values below 500~Kbps are considered a service outage.} 
	\label{figUL}
\end{figure}

The second of the network layer parameters analyzed is throughput. This analysis includes both downlink and uplink throughput for single-connectivity and multi-connectivity strategies (see Section III.C). Table~IV and Figs.~\ref{figDL} and \ref{figUL} show the statistics and CDF distributions acquired along the route.

\subsubsection{Downlink throughput}

Focusing on the downlink, an outage probability of 7.4\%, i.e., P(DL $<$ 500 Kbps), is observed for cellular operator A, and 4.5\% for the satellite link. However, considering the median values, they are able to fulfill the 100 Mbps target data rate at least with $50\%$ probability. In other words, these two networks seem capable of reaching the target when they are not in service outage. On the other hand, this outage probability decreases to 1.3\% for operator~B. However, despite having a low service outage probability compared to the other networks, operator B has a median downlink value of 86.2 Mbps. This means that the likelihood of not establishing a stable connection is low compared to the other operators, but the throughput once the connection is established may not meet the high bandwidth target. This is illustrated in the CDF for operator B in Fig. \ref{figDL}, where the CDF distribution shows a more linear trend. Regarding the tails, Table~IV also denotes the throughput for the worst 10\%, 5\%, and 1\% of cases evaluated. The comparison between cellular and satellite networks shows better performance of the satellite link with minimum throughput of 77.7 Mbps in the worst $10\%$ scenario, resulting in significantly better performance compared to both cellular operators, where the best telecom provider case can offer 12.9~Mbps for the worst 10\% scenario. Compared to SC strategies, MC approaches achieve a service outage of only 0.7\% and 0.5\% for cellular-based MC and cellular-satellite MC, respectively. The median values satisfactorily reach the established throughput target. In the distribution tails, the cellular-satellite multi-connectivity solution stands out as it achieves values above 90 Mbps for the worst 10\% and 5\% of cases, dropping to approximately 30 Mbps for the worst 1\% of samples. Thus, the integration of these technologies significantly improves the downlink in these rural environments.

\subsubsection{Uplink throughput}

In the uplink analysis, the satellite link exhibits notably low throughput. Specifically, as shown in Fig. \ref{figUL}, there is a high percentage of satellite samples clustered around 20 Mbps, which differs significantly from the downlink. This discrepancy may be due to Starlink being primarily optimized for downlink, either due to (i) resource allocation in the time scheme or (ii) limitations in user hardware. Note that the effective isotropic radiated power (EIRP) in the user equipment is not comparable to the gains achievable for satellites and terrestrial gateways, so the path loss in the uplink is more constrained than in the downlink, requiring more robust constellations due to the SNR and dynamic margin available in the link budget. Consequently, not only does the throughput decrease, but the service outage i.e., P(UL $<$ 500 Kbps), also increases to 9.8\%. This reduced throughput performance is also observed in cellular operators, with median values around 50 Mbps. In any case, the comparison with MC approaches still offers some benefits over standard connectivity strategies. For example, the CDF of the cellular-satellite MC case reports an improvement of a few Mbps compared to the connectivity case with operator A, mainly due to the low values found for Starlink. However, in the low throughput regime, the robustness provided by both interfaces ensures a minimum uplink of 23.9 Mbps in the worst 10\% of cases compared to single connectivity (9.9~Mbps and 1.2~Mbps). For values above 25 Mbps, cellular-based MC provides the best results, with improvements of around 20 Mbps compared to the SC cases.

\renewcommand{\arraystretch}{1.5}
\setlength{\tabcolsep}{4pt}

\begin{table*}[t]
\centering
\caption{Throughput statistics for single-connectivity and multi-connectivity approaches in rural scenarios}
\label{table3}
\resizebox{\textwidth}{!}{%
\begin{tabular}{|c||ccccccc||ccccccc|}
\hline
\hline
 &
  \multicolumn{7}{c||}{DL (Mbps)} &
  \multicolumn{7}{c|}{UL (Mbps)} \\ \hline \hline
 &
  \multicolumn{1}{c|}{Median} &
  \multicolumn{1}{c|}{Mean} &
  \multicolumn{1}{c|}{Std. dev} &
  \multicolumn{1}{c|}{Outage} &
  \multicolumn{1}{c|}{10\%} &
  \multicolumn{1}{c|}{5\%} &
  1\% &
  \multicolumn{1}{c|}{Median} &
  \multicolumn{1}{c|}{Mean} &
  \multicolumn{1}{c|}{Std. dev} &
  \multicolumn{1}{c|}{Outage} &
  \multicolumn{1}{c|}{10\%} &
  \multicolumn{1}{c|}{5\%} &
  1\% \\ \hline \hline
Operator A &
  \multicolumn{1}{c|}{98.4} &
  \multicolumn{1}{c|}{82.3} &
  \multicolumn{1}{c|}{29.4} &
  \multicolumn{1}{c|}{7.4\%} &
  \multicolumn{1}{c|}{1.5} &
  \multicolumn{1}{c|}{0} &
  0 &
  \multicolumn{1}{c|}{50.3} &
  \multicolumn{1}{c|}{52} &
  \multicolumn{1}{c|}{26.5} &
  \multicolumn{1}{c|}{3.3\%} &
  \multicolumn{1}{c|}{9.9} &
  \multicolumn{1}{c|}{0.4} &
  0 \\ 
Operator B &
  \multicolumn{1}{c|}{86.2} &
  \multicolumn{1}{c|}{59.3} &
  \multicolumn{1}{c|}{33.8} &
  \multicolumn{1}{c|}{1.3\%} &
  \multicolumn{1}{c|}{12.9} &
  \multicolumn{1}{c|}{6.9} &
  0 &
  \multicolumn{1}{c|}{45.5} &
  \multicolumn{1}{c|}{48.5} &
  \multicolumn{1}{c|}{32.7} &
  \multicolumn{1}{c|}{1.5\%} &
  \multicolumn{1}{c|}{3.0} &
  \multicolumn{1}{c|}{1.1} &
  0 \\ 
Satellite &
  \multicolumn{1}{c|}{99.9} &
  \multicolumn{1}{c|}{92.5} &
  \multicolumn{1}{c|}{19} &
  \multicolumn{1}{c|}{4.5\%} &
  \multicolumn{1}{c|}{77.7} &
  \multicolumn{1}{c|}{5.8} &
  0 &
  \multicolumn{1}{c|}{21.8} &
  \multicolumn{1}{c|}{24.6} &
  \multicolumn{1}{c|}{11.6} &
  \multicolumn{1}{c|}{9.8\%} &
  \multicolumn{1}{c|}{1.2} &
  \multicolumn{1}{c|}{0} &
  0 \\ \hline 
Cellular-based MC &
  \multicolumn{1}{c|}{99.7} &
  \multicolumn{1}{c|}{87.7} &
  \multicolumn{1}{c|}{21} &
  \multicolumn{1}{c|}{0.7\%} &
  \multicolumn{1}{c|}{68.7} &
  \multicolumn{1}{c|}{42.5} &
  7.3 &
  \multicolumn{1}{c|}{68.3} &
  \multicolumn{1}{c|}{62.6} &
  \multicolumn{1}{c|}{26.5} &
  \multicolumn{1}{c|}{0.6\%} &
  \multicolumn{1}{c|}{22.5} &
  \multicolumn{1}{c|}{12.3} &
  1.2 \\ 
Cellular-Satellite MC &
  \multicolumn{1}{c|}{100} &
  \multicolumn{1}{c|}{96.4} &
  \multicolumn{1}{c|}{11.5} &
  \multicolumn{1}{c|}{0.5\%} &
  \multicolumn{1}{c|}{98.1} &
  \multicolumn{1}{c|}{92.7} &
  32.7 &
  \multicolumn{1}{c|}{55.8} &
  \multicolumn{1}{c|}{56.7} &
  \multicolumn{1}{c|}{24.2} &
  \multicolumn{1}{c|}{1.0\%} &
  \multicolumn{1}{c|}{23.9} &
  \multicolumn{1}{c|}{20.9} &
  0 \\ \hline \hline
\end{tabular}%
}
\end{table*}


In summary, the results presented in Section IV.C demonstrate that a multi-connectivity strategy combining cellular and satellite links is capable of meeting a target of 100 Mbps for downlink with service outages in only 0.5\% of the samples studied. As illustrated in Table~IV, a considerable improvement in service stability is achieved, as indicated by the tails of the CDF. In the case of the uplink, the benefits of multi-connectivity with a satellite link are not as obvious due to the low throughput offered by the satellite compared to cellular networks. However, low throughput regimes show an increase in network robustness and reliability with service outages below 1.0\%, as well as improvements of around 20 Mbps when using cellular-based multi-connectivity compared to single-connectivity approaches.

\section{Discussion of the results from a rural IoT use case point of view}

To emphasize the benefits provided by multi-connectivity strategies, a quantitative analysis is conducted to evaluate how each connectivity strategy can meet the KPIs of different types of connectivity and use cases based on rural environments. The measurement scenario proposed throughout the study aims to resemble conditions similar to those faced by the aforementioned use cases. In this way, the success KPI availability serves as an indicator of the performance between single-connectivity and multi-connectivity strategies. Therefore, Table~\ref{tab:comparison} presents the reliability with which the latency and downlink/uplink throughput KPIs are met for each type of connectivity solution given the use cases identified in Section II (see Appendix A for confidence intervals). Those percentages marked in bold indicate those KPIs given the use case and technology for which the required minimum service availability, shown in Table~I, is met. Additionally, the yellow cells highlight those requirements that, although not fully met, are within less than 1\% of the required availability.

Observing the KPI satisfaction rate for single-connectivity cases in cellular networks, it can be seen that only the minimum criterion is met in use cases where the minimum availability is $90\%$ ($10^{-2}$). These results align with those shown in Figs. 7, 8, and 9, where the tails in cellular networks generally do not allow for the fulfillment of services with minimum availabilities exceeding $90\%$. In the case of satellite single-connectivity, this technology is able to meet a larger number of use cases in terms of latency due to the stability of the link given low jitter values described in Section~IV.B. On the other hand, in terms of downlink throughput, only those cases with a minimum availability of $90\%$ can be satisfied. Regarding uplink throughput, satellite technology is unable to meet the minimum KPI availability due to the satellite optimization for downlink previously discussed in Section~IV.C.

For the multi-connectivity approaches, the improvement in cellular multi-connectivity compared to single connectivity cases is noteworthy, due to the redundancy and diversity gain provided by having available two cellular networks. Additionally, the case of cellular and satellite multi-connectivity is remarkable since the minimum reliability values are reached for most cases in terms of latency and downlink data rate. However, it can also be noted that in terms of uplink throughput, these solutions are still insufficient to meet the minimum requirements. Nevertheless, from a qualitative perspective, the network availability for the required uplink throughput values generally increases above $90\%$ for cellular multi-connectivity and above $95\%$ for cellular-satellite multi-connectivity.

\renewcommand{\arraystretch}{1.5}
\setlength{\tabcolsep}{4pt}

\begin{table*}[!t]
\caption{Comparison of the different technologies for the KPIs of each use case based on the availability of latencies, downlink and uplink data rates required}
\label{tab:comparison}
\resizebox{\textwidth}{!}{%
\begin{tabular}{|c||ccc||ccc||ccc||ccc||ccc|}
\hline\hline
\multirow{2}{*}{Type of connectivity/Use Case} &
  \multicolumn{3}{c||}{\thead{SC - Operator A \\ 4G/5G}} &
  \multicolumn{3}{c||}{\thead{SC - Operator B \\ 4G/5G}} &
  \multicolumn{3}{c||}{\thead{SC - Satellite \\ Starlink}} &
  \multicolumn{3}{c||}{\thead{MC - Cellular \\ 4G/5G and 4G/5G}} &
  \multicolumn{3}{c|}{\thead{MC - Cellular Satellite \\ 4G/5G and Starlink}} \\ \cline{2-16}
 &
  \multicolumn{1}{c|}{Lat} &
  \multicolumn{1}{c|}{DL} &
  UL &
  \multicolumn{1}{c|}{Lat} &
  \multicolumn{1}{c|}{DL} &
  UL &
  \multicolumn{1}{c|}{Lat} &
  \multicolumn{1}{c|}{DL} &
  UL &
  \multicolumn{1}{c|}{Lat} &
  \multicolumn{1}{c|}{DL} &
  UL &
  \multicolumn{1}{c|}{Lat} &
  \multicolumn{1}{c|}{DL} &
  UL \\ \hline\hline
M2DC &
  \multicolumn{1}{c|}{88.1\%} &
  \multicolumn{1}{c|}{80.5\%} &
  50.4\% &
  \multicolumn{1}{c|}{84.4\%} &
  \multicolumn{1}{c|}{67.3\%} &
  46.8\% &
  \multicolumn{1}{c|}{77.3\%} &
  \multicolumn{1}{c|}{92.4\%} &
  5.1\% &
  \multicolumn{1}{c|}{97.6\%} &
  \multicolumn{1}{c|}{93.9\%} &
  67.8\% &
  \multicolumn{1}{c|}{\cellcolor{yellowcell}98.2\%} &
  \multicolumn{1}{c|}{\cellcolor{yellowcell}98.5\%} &
  58.3\% \\ \hline
M2SD &
  \multicolumn{1}{c|}{88.1\%} &
  \multicolumn{1}{c|}{87.3\%} &
  90.0\% &
  \multicolumn{1}{c|}{84.4\%} &
  \multicolumn{1}{c|}{92.4\%} &
  79.3\% &
  \multicolumn{1}{c|}{77.3\%} &
  \multicolumn{1}{c|}{94.6\%} &
  86.4\% &
  \multicolumn{1}{c|}{97.6\%} &
  \multicolumn{1}{c|}{\cellcolor{yellowcell}98.9\%} &
  96.3\% &
  \multicolumn{1}{c|}{\cellcolor{yellowcell}98.2\%} &
  \multicolumn{1}{c|}{\cellcolor{greencell}\textbf{99.3\%}} &
  \cellcolor{yellowcell}98.7\% \\ \hline
M2M &
  \multicolumn{1}{c|}{88.1\%} &
  \multicolumn{1}{c|}{88.4\%} &
  92.4\% &
  \multicolumn{1}{c|}{84.4\%} &
  \multicolumn{1}{c|}{96.1\%} &
  86.1\% &
  \multicolumn{1}{c|}{77.3\%} &
  \multicolumn{1}{c|}{95.1\%} &
  88.4\% &
  \multicolumn{1}{c|}{97.6\%} &
  \multicolumn{1}{c|}{\cellcolor{greencell}\textbf{99.2\%}} &
  97.9\% &
  \multicolumn{1}{c|}{\cellcolor{yellowcell}98.2\%} &
  \multicolumn{1}{c|}{\cellcolor{greencell}\textbf{99.4\%}} &
  \cellcolor{yellowcell}98.9\% \\ \hline\hline\hline
UC1 (\textit{Precision Agriculture})&
  \multicolumn{1}{c|}{90.6\%} &
  \multicolumn{1}{c|}{88.4\%} &
  92.4\% &
  \multicolumn{1}{c|}{91.5\%} &
  \multicolumn{1}{c|}{96.1\%} &
  86.1\% &
  \multicolumn{1}{c|}{\cellcolor{greencell}\textbf{99.8\%}} &
  \multicolumn{1}{c|}{95.1\%} &
  88.4\% &
  \multicolumn{1}{c|}{\cellcolor{yellowcell}98.6\%} &
  \multicolumn{1}{c|}{\cellcolor{greencell}\textbf{99.2\%}} &
  97.9\% &
  \multicolumn{1}{c|}{\cellcolor{greencell}\textbf{99.9\%}} &
  \multicolumn{1}{c|}{\cellcolor{greencell}\textbf{99.4\%}} &
  \cellcolor{yellowcell}98.9\% \\ \hline
UC2 (\textit{Digital Twin})&
  \multicolumn{1}{c|}{90.6\%} &
  \multicolumn{1}{c|}{90.8\%} &
  92.4\% &
  \multicolumn{1}{c|}{91.5\%} &
  \multicolumn{1}{c|}{\cellcolor{yellowcell}98.0\%} &
  86.1\% &
  \multicolumn{1}{c|}{\cellcolor{greencell}\textbf{99.8\%}} &
  \multicolumn{1}{c|}{95.5\%} &
  88.4\% &
  \multicolumn{1}{c|}{\cellcolor{yellowcell}98.6\%} &
  \multicolumn{1}{c|}{\cellcolor{greencell}\textbf{99.2\%}} &
  97.9\% &
  \multicolumn{1}{c|}{\cellcolor{greencell}\textbf{99.9\%}} &
  \multicolumn{1}{c|}{\cellcolor{greencell}\textbf{99.5\%}} &
  \cellcolor{yellowcell}98.9\% \\ \hline
UC3 (\textit{Remote Assistance})&
  \multicolumn{1}{c|}{88.1\%} &
  \multicolumn{1}{c|}{90.8\%} &
  82.6\% &
  \multicolumn{1}{c|}{84.4\%} &
  \multicolumn{1}{c|}{98.0\%} &
  69.2\% &
  \multicolumn{1}{c|}{77.3\%} &
  \multicolumn{1}{c|}{95.5\%} &
  60.7\% &
  \multicolumn{1}{c|}{97.6\%} &
  \multicolumn{1}{c|}{\cellcolor{yellowcell}99.2\%} &
  91.6\% &
  \multicolumn{1}{c|}{98.2\%} &
  \multicolumn{1}{c|}{\cellcolor{yellowcell}99.5\%} &
  95.6\% \\ \hline
UC4 (\textit{Emergency Response})&
  \multicolumn{1}{c|}{88.1\%} &
  \multicolumn{1}{c|}{90.8\%} &
  82.6\% &
  \multicolumn{1}{c|}{84.4\%} &
  \multicolumn{1}{c|}{98.0\%} &
  69.2\% &
  \multicolumn{1}{c|}{77.3\%} &
  \multicolumn{1}{c|}{95.5\%} &
  60.7\% &
  \multicolumn{1}{c|}{97.6\%} &
  \multicolumn{1}{c|}{\cellcolor{yellowcell}99.2\%} &
  91.6\% &
  \multicolumn{1}{c|}{98.2\%} &
  \multicolumn{1}{c|}{\cellcolor{yellowcell}99.5\%} &
  95.6\% \\ \hline
UC5 (\textit{Livestock Logistics})&
  \multicolumn{1}{c|}{90.6\%} &
  \multicolumn{1}{c|}{86.4\%} &
  93.1\% &
  \multicolumn{1}{c|}{91.5\%} &
  \multicolumn{1}{c|}{89.3\%} &
  95.4\% &
  \multicolumn{1}{c|}{\cellcolor{yellowcell}99.8\%} &
  \multicolumn{1}{c|}{94.5\%} &
  90.2\% &
  \multicolumn{1}{c|}{98.6\%} &
  \multicolumn{1}{c|}{98.5\%} &
  \cellcolor{yellowcell}99.1\% &
  \multicolumn{1}{c|}{\cellcolor{greencell}\textbf{99.9\%}} &
  \multicolumn{1}{c|}{\cellcolor{yellowcell}99.3\%} &
  \cellcolor{yellowcell}98.9\% \\ \hline
UC6 (\textit{Automated Access})&
  \multicolumn{1}{c|}{94.6\%} &
  \multicolumn{1}{c|}{90.8\%} &
  92.4\% &
  \multicolumn{1}{c|}{\cellcolor{yellowcell}98.5\%} &
  \multicolumn{1}{c|}{\cellcolor{yellowcell}98.0\%} &
  86.1\% &
  \multicolumn{1}{c|}{\cellcolor{greencell}\textbf{99.9\%}} &
  \multicolumn{1}{c|}{95.5\%} &
  88.4\% &
  \multicolumn{1}{c|}{\cellcolor{greencell}\textbf{99.6\%}} &
  \multicolumn{1}{c|}{\cellcolor{greencell}\textbf{99.2\%}} &
  97.9\% &
  \multicolumn{1}{c|}{\cellcolor{greencell}\textbf{99.9\%}} &
  \multicolumn{1}{c|}{\cellcolor{greencell}\textbf{99.5\%}} &
  \cellcolor{yellowcell}98.9\% \\ \hline
UC7 (\textit{Computer Vision})&
  \multicolumn{1}{c|}{94.6\%} &
  \multicolumn{1}{c|}{88.4\%} &
  90.0\% &
  \multicolumn{1}{c|}{\cellcolor{yellowcell}98.5\%} &
  \multicolumn{1}{c|}{96.1\%} &
  79.3\% &
  \multicolumn{1}{c|}{\cellcolor{greencell}\textbf{99.9\%}} &
  \multicolumn{1}{c|}{95.1\%} &
  86.4\% &
  \multicolumn{1}{c|}{\cellcolor{greencell}\textbf{99.6\%}} &
  \multicolumn{1}{c|}{\cellcolor{greencell}\textbf{99.2\%}} &
  96.3\% &
  \multicolumn{1}{c|}{\cellcolor{greencell}\textbf{99.9\%}} &
  \multicolumn{1}{c|}{\cellcolor{greencell}\textbf{99.4\%}} &
  \cellcolor{yellowcell}98.7\% \\ \hline
UC8 (\textit{Predictive Analytics})&
  \multicolumn{1}{c|}{\cellcolor{greencell}\textbf{90.6\%}} &
  \multicolumn{1}{c|}{88.4\%} &
  \cellcolor{greencell}\textbf{92.4\%} &
  \multicolumn{1}{c|}{\cellcolor{greencell}\textbf{91.5\%}} &
  \multicolumn{1}{c|}{\cellcolor{greencell}\textbf{96.1\%}} &
  86.1\% &
  \multicolumn{1}{c|}{\cellcolor{greencell}\textbf{99.8\%}} &
  \multicolumn{1}{c|}{\cellcolor{greencell}\textbf{95.1\%}} &
  88.4\% &
  \multicolumn{1}{c|}{\cellcolor{greencell}\textbf{98.6\%}} &
  \multicolumn{1}{c|}{\cellcolor{greencell}\textbf{99.2\%}} &
  \cellcolor{greencell}\textbf{97.9\%} &
  \multicolumn{1}{c|}{\cellcolor{greencell}\textbf{99.9\%}} &
  \multicolumn{1}{c|}{\cellcolor{greencell}\textbf{99.4\%}} &
  \cellcolor{greencell}\textbf{98.9\%} \\ \hline
UC9 (\textit{Remote Sensing})&
  \multicolumn{1}{c|}{\cellcolor{greencell}\textbf{90.6\%}} &
  \multicolumn{1}{c|}{\cellcolor{greencell}\textbf{90.8\%}} &
  \cellcolor{greencell}\textbf{92.4\%} &
  \multicolumn{1}{c|}{\cellcolor{greencell}\textbf{91.5\%}} &
  \multicolumn{1}{c|}{\cellcolor{greencell}\textbf{98.0\%}} &
  86.1\% &
  \multicolumn{1}{c|}{\cellcolor{greencell}\textbf{99.8\%}} &
  \multicolumn{1}{c|}{\cellcolor{greencell}\textbf{95.5\%}} &
  88.4\% &
  \multicolumn{1}{c|}{\cellcolor{greencell}\textbf{98.6\%}} &
  \multicolumn{1}{c|}{\cellcolor{greencell}\textbf{99.2\%}} &
  \cellcolor{greencell}\textbf{97.9\%} &
  \multicolumn{1}{c|}{\cellcolor{greencell}\textbf{99.9\%}} &
  \multicolumn{1}{c|}{\cellcolor{greencell}\textbf{99.5\%}} &
  \cellcolor{greencell}\textbf{98.9\%} \\ \hline\hline
Technology-ready use cases &
  \multicolumn{1}{c|}{2/9} &
  \multicolumn{1}{c|}{1/9} &
  2/9 &
  \multicolumn{1}{c|}{2/9} &
  \multicolumn{1}{c|}{2/9} &
  0/9 &
  \multicolumn{1}{c|}{6/9} &
  \multicolumn{1}{c|}{2/9} &
  0/9 &
  \multicolumn{1}{c|}{4/9} &
  \multicolumn{1}{c|}{6/9} &
  2/9 &
  \multicolumn{1}{c|}{7/9} &
  \multicolumn{1}{c|}{6/9} &
  2/9 \\ \hline\hline
\end{tabular}%
}
\end{table*}

From an individual use case perspective, satellite-cellular multi-connectivity has empirically proven to be the most promising in terms of performance. In particular, the UC1 and UC2 cases, related to monitoring and digital twin in vineyards, require latencies below 400 ms and approximately 5 Mbps symmetrical bandwidths with reliabilities above 99\%. These requirements are satisfactorily met in the case of latency and downlink througput with reliabilities exceeding 99.9\% and 99.4\% respectively. In the downlink case, the minimum requirement is met in 98.9\% of the cases, virtually at the border of the required reliability. As is also the case for the other use cases, the uplink is usually the most constrained KPI. A possible improvement on this indicator could be through further densification of the network to improve SNR conditions in the channel, or a larger allocation of resources in TDD/FDD schemes to the uplink channel. Use cases related to remote operational support and complex situation awareness service in forests (UC3 and UC4) demand latencies of less than 100 ms and downlink/uplink data rates of 1/20 Mbps. Particularly notable is the requirement for uplink due to the associated need for video transmission. Additionally, due to the critical nature of the use cases, the required reliability increases to 99.9\%. While latency and downlink availability reach values relatively close to the requirement, i.e., 98.2\% and 99.5\%, respectively, the uplink availability is noticeably lower at 95.6\%, highlighting the need for improved infrastructure or deployment strategies even when applying multi-connectivity. UC5 (monitoring of livestock transport along routes) availability is 99.9\% for requested latencies below 400~ms and 99.3\%/98.9\% for 14/1 (DL/UL) Mbps data rates. Similarly, the use cases related to livestock load monitoring and truck license plate recognition (UC6 and UC7) meet the latency, downlink, and uplink requirements with reliabilities of at least 99.9\%, 99.4\% and 98.7\%. Finally, the use cases UC8 and UC9, early disease and monitoring of pest insect traps in olive crops are currently fully functional with the technology tested in this work due to KPI availabilities of 99.9\%, 99.5\% and 98.9\% given a minimum required reliability of 90\%. As a general note, the use of cellular-satellite multi-connectivity seems to be sufficient to meet the required use cases KPIs in terms of latency (7 out of 9 use cases) and downlink (6 out of 9 use cases). However, the current limitation is in uplink (2 out of 9 use cases), where despite finding availabilities above 95\%, in some cases within less than 1\% of the required availability (see yellow cells in Table~V), they are still far from 99\% or 99.9\% expected reliabilities. The limitations in the uplink are mainly due to Starlink's asymmetric system design, which prioritizes downlink performance. This is mainly due to the smaller size of the user terminal on the ground, which implies: (i) limited transmission power, (ii) smaller effective antenna aperture, and (iii) reduced beam reconfigurability due to the limited size of the phased-array antenna. Additionally, more bandwidth is allocated to the downlink than to the uplink, further increasing the asymmetry in achievable data rates. Therefore, it should be noted that network operators' efforts and infrastructure deployment should be focused in this direction, implementing enhancements at the terminal or network level, such as higher gain user terminals or scheduling policies to improve uplink performance in satellite systems.

In summary, the availability and reliability values of the various KPIs in Table~V show a significant improvement compared to single-connectivity approaches, as well as an initial step toward enabling seamless connectivity services in rural environments. Thus, a deployment based on multi-connectivity strategies is recommended when a network with high availability, low latency, and minimum throughput requirements is needed. Both TN-TN (5G-5G) and \mbox{TN-NTN (5G-Sat.)} multi-connectivity solutions demonstrate superior effectiveness and efficiency for the proposed use cases compared to traditional terrestrial network solutions under single-connectivity conditions. As indicated in Section III.B, the multi-connectivity results presented in this work are based on a full duplication strategy, where all packets are duplicated regardless of network conditions. This approach is intended to determine the maximum performance achievable by multi-connectivity strategies. Future implementations may exploit load balancing or dynamic switching based on smart packet duplication given real-time link quality estimation. For instance, using parameters such as RSRP to perform predictive switching between available interfaces. This could result in packet duplication when the quality of the primary interface degrades, or dynamic switching between interfaces, to always transmit through the most reliable link. The main advantage of these strategies lies in reducing network resource usage, though at the cost of requiring accurate and optimal predictions of interface performance.

\subsection{Cost and power consumption of the proposed solutions}

\begin{table*}[!t]
\centering
\caption{Summary of the cost and power consumption of each of the single-connectivity and multi-connectivity solutions proposed for rural areas}
\label{tab:cost_power}
\resizebox{\textwidth}{!}{%
\begin{tabular}{|c||c||c||c||c|}
\hline \hline
Type of connectivity &
  \begin{tabular}[c]{@{}c@{}}SC - Cellular          \\ 4G/5G\end{tabular} &
  \begin{tabular}[c]{@{}c@{}}SC - Satellite         \\ Starlink\end{tabular} &
  \begin{tabular}[c]{@{}c@{}}MC - Cellular          \\ 4G/5G and 4G/5G\end{tabular} &
  \begin{tabular}[c]{@{}c@{}}MC - Cellular Satellite\\ 4G/5G and Starlink\end{tabular} \\ \hline \hline
Cost &
  \begin{tabular}[c]{@{}c@{}} $\sim$200\$ - 250\$ (5G Modem)                                     \\ $\sim$10\$ - 20\$ (5G Data Plan)                                    \\ Total: $\sim$225\$ + $\sim$15\$/month\end{tabular} &
  \begin{tabular}[c]{@{}c@{}} $\sim$250\$ - 300\$ (Dish Antenna)                                 \\ $\sim$40\$ - 50\$ (Sat. Data Plan)                                  \\ Total: $\sim$275\$ + $\sim$45\$/month\end{tabular} &
  \begin{tabular}[c]{@{}c@{}} 2x $\sim$200\$ - 250\$ (5G Modem)                                  \\ 2x $\sim$10\$ - 20\$ (5G Data Plan)                                 \\ Total: $\sim$450\$ + $\sim$30\$/month\end{tabular} &
  \begin{tabular}[c]{@{}c@{}} $\sim$200\$ - 250\$ (5G Modem) / $\sim$250\$ - 300\$ (Dish Antenna)\\ $\sim$10\$ - 20\$(5G Data Plan) / $\sim$40\$ - 50\$ (Sat. Data Plan)\\ Total: $\sim$500\$ + $\sim$60\$/month\end{tabular} \\ \hline
Power Consumption &
  $\sim$7W (5G Modem) &
  $\sim$75W (Dish Antenna) &
  $\sim$14W (2x 5G Modem) &
  $\sim$82W (5G Modem and Dish Antenna) \\ \hline \hline
\end{tabular}%
}
\end{table*}

To conclude this Section, Table~VI presents a summary of the estimated cost and power consumption for each type of connectivity solution. These factors are considered due to their importance in rural connectivity deployments. The results focus on the two types of single-connectivity solutions and the two multi-connectivity solutions.

Regarding the cost, the price has been divided into the required hardware equipment (5G modem or dish antenna) and the monthly subscription to cellular or satellite services. In terms of hardware, the price range is similar, while there is a notable difference in the monthly service subscription, with Starlink being more expensive. This is expected, as satellite technology is still relatively new and under active development, and a price reduction can be anticipated as its market penetration increases. Note that the prices refer to the specific equipment used for the measurement campaign in this work, presented in Section III.B. In terms of power consumption and according to the manufacturers' datasheets, the 5G modem and dish antenna consume approximately 7W and 75W, respectively, during periods of high data rate exchange. This implies an order of magnitude higher for consumption in the case of the satellite system, which is expected due to the larger antenna size and the beamforming requirements needed to compensate for the long link distances. It is also worth noting that power consumption during idle states or low data rates is expected to be significantly lower for both technologies.

Finally, it should be noted that the above analysis refers exclusively to 5G or satellite connectivity solutions. For actual implementation, a small computer is required, such as the GW6400 used in this work, or a NUC that allows the different interfaces to be interconnected. This equipment will determine the scalability of the hardware based on the number of interfaces available. In the case of the GW6400, it has four Mini-PCIe ports and five Ethernet ports, enabling up to nine interfaces. Regarding the \textit{mpconn} duplication tool software~\cite{mpconn}, it has been released as open-source software and is freely available for use.

\section{Conclusions}

This work evaluates the feasibility of operating rural IoT use cases requiring different network communication capabilities. To do so, a measurement campaign is performed to experimentally evaluate the benefits of integration between several cellular networks, and also cellular and satellite networks. Specifically, it addresses the use case based on rural scenarios where the penetration of mobile networks is generally lower than their urban counterparts. Thus, a testbed is developed and integrated into a car based on satellite connectivity solutions, through Starlink, and cellular connectivity, through two cellular operators. Additionally, a packet duplication tool at the network layer is developed for testing multi-connectivity strategies.

The study evaluates the feasibility of operating key rural applications under different connectivity scenarios by comparing measured KPIs, such as latency, jitter, and throughput, against their requirements. Results show that single-connectivity solutions (5G TN or satellite NTN) are often insufficient to support demanding use cases like autonomous farming equipment or real-time livestock monitoring, due to high latencies and unstable throughput. Multi-connectivity strategies, particularly those combining terrestrial and satellite links, significantly improve performance, achieving stable latencies below 100 ms and maintaining stable uplink/downlink rates, thus enabling reliable operation of use cases requiring high data rates or low delay. For instance, TN-NTN (5G-Sat.) configurations meet KPI thresholds at least 98\% of the time for latency, 99\% for downlink, and 95\% for uplink, supporting applications ranging from precision agriculture to environmental monitoring and forest management.

Given the benefits obtained from integrating both technologies, future development of integrated systems considering multiple communication technologies is promising. While specific performance metrics may vary in other geographies since this study is based on a specific rural deployment, the insights and multi-connectivity gains explained throughout our work are expected to be valid in other rural scenarios characterized by sparse population, limited land infrastructure, intermittent coverage, and traffic patterns typical of rural contexts. Thus, these findings are generalizable to similar environments where cellular networks are available but present coverage gaps, and satellite connectivity can act as a complementary solution. In this study, every packet in the network layer has been duplicated to assess the raw benefit achievable through the integration of multiple networks and technologies. Future research lines include smart packet duplication, where a prediction of the links to be employed at any given time is performed based on KPIs of the physical layer, such as the signal strength. Thus, this strategy might be potentially beneficial from a resource-saving perspective.

\appendices

\section{Confidence intervals $-$ DKW inequality and Wilson score interval}

\renewcommand{\arraystretch}{1.5}
\setlength{\tabcolsep}{4pt}

\begin{table*}[!b]
\caption{DKW inequality - Confidence intervals for throughput CDFs and latency CCDFs with $\alpha = 0.01$}
\label{tab:comparison_confidence_interval}
\resizebox{\textwidth}{!}{%
\begin{tabular}{|c||ccc||ccc||ccc||ccc||ccc|}
\hline\hline
\multirow{2}{*}{Type of connectivity} &
  \multicolumn{3}{c||}{\thead{SC - Operator A \\ 4G/5G}} &
  \multicolumn{3}{c||}{\thead{SC - Operator B \\ 4G/5G}} &
  \multicolumn{3}{c||}{\thead{SC - Satellite \\ Starlink}} &
  \multicolumn{3}{c||}{\thead{MC - Cellular \\ 4G/5G and 4G/5G}} &
  \multicolumn{3}{c|}{\thead{MC - Cellular Satellite \\ 4G/5G and Starlink}} \\ \cline{2-16}
 &
  \multicolumn{1}{c|}{Lat} &
  \multicolumn{1}{c|}{DL} &
  UL &
  \multicolumn{1}{c|}{Lat} &
  \multicolumn{1}{c|}{DL} &
  UL &
  \multicolumn{1}{c|}{Lat} &
  \multicolumn{1}{c|}{DL} &
  UL &
  \multicolumn{1}{c|}{Lat} &
  \multicolumn{1}{c|}{DL} &
  UL &
  \multicolumn{1}{c|}{Lat} &
  \multicolumn{1}{c|}{DL} &
  UL \\ \hline\hline
Confidence Interval ($\varepsilon_{DKW}$) &
  \multicolumn{1}{c|}{$6.7 \times 10^{-3}$} &
  \multicolumn{1}{c|}{$2.7 \times 10^{-4}$} &
  $3.1 \times 10^{-4}$ &
  \multicolumn{1}{c|}{$6.4 \times 10^{-3}$} &
  \multicolumn{1}{c|}{$2.7 \times 10^{-4}$} &
  $3.3 \times 10^{-4}$ &
  \multicolumn{1}{c|}{$6.8 \times 10^{-3}$} &
  \multicolumn{1}{c|}{$2.7 \times 10^{-4}$} &
  $2.4 \times 10^{-4}$ &
  \multicolumn{1}{c|}{$6.3 \times 10^{-3}$} &
  \multicolumn{1}{c|}{$2.4 \times 10^{-4}$} &
  $2.9 \times 10^{-4}$ &
  \multicolumn{1}{c|}{$7.1 \times 10^{-3}$} &
  \multicolumn{1}{c|}{$2.3 \times 10^{-4}$} &
  $3.6 \times 10^{-4}$ \\ \hline\hline
\end{tabular}%
}
\end{table*}

This Appendix A presents the confidence intervals for the latency CCDF (Fig. 7) and downlink and uplink throughput CDFs (Figs. 8 and 9) with the aim of demonstrating the statistical reliability of the results presented throughout the work. The analysis of the CCDF and CDFs is based on statistics from \textit{ping} packets for latency and UDP packets for throughput.

Given a sample of size $n$, $(x_1,x_2,...,x_{n-1}, x_n)$, an empirical CDF can be defined as

\begin{equation}
\hat{F_n}(x)=\frac{1}{n} \sum_{i=1}^n 1_{\left\{X_i \leq x\right\}}
\end{equation}

Based on the Dvoretzky–Kiefer–Wolfowitz (DKW) inequality~\cite{DKW}, a bound can be generated between the CDF F(x) and the empirical CDF $\hat{F_n}(x)$ on the worst case distance $\varepsilon_{DKW}$ between both as follows

\begin{equation}
P\left(\sup _x|F(x)-\hat{F_n}(x)|>\varepsilon_{DKW}\right) \leq 2 e^{-2 n \varepsilon_{DKW}^2}
\end{equation}

Thus, $\varepsilon_{DKW}$ can be considered as an upper and lower confidence interval for the entire range of the CDF, i.e., $\hat{F_n}(x) \pm \varepsilon_{DKW}$. Therefore, $\varepsilon_{DKW}$ is calculated as

\begin{equation}
\varepsilon_{DKW}=\sqrt{\frac{1}{2 n} \ln \left(\frac{2}{\alpha}\right)},
\end{equation}

where the confidence interval contains $F(x)$ with probability $1-\alpha$.

Given the number of samples $n$ from our empirical CDFs and an $\alpha = 0.01$, Table~\ref{tab:comparison_confidence_interval} shows the confidence intervals for the throughput CDFs and latency CCDFs given the different type of connectivity available during the measurement campaign. In the case of downlink and uplink, $\varepsilon_{DKW}$ is in the order of $10^{-4}$. This implies a confidence interval of around $\pm$0.01\%, which is sufficient even to validate empirical CDFs for use cases with 99.9\% reliability. In the case of latency, the confidence interval order is $10^{-3}$, which implies a confidence interval of $\pm$0.1\%. While this interval may be valid for use cases with reliability requirements of 90\% and 99\%, it may be insufficient for those cases with reliabilities of 99.9\%. As indicated at the beginning of the Appendix, the Dvoretzky–Kiefer–Wolfowitz inequality establishes an interval for the entire range of the CDF, which is a conservative approximation that limits the accuracy of the confidence interval at tails of the distribution, e.g., at 99.9\%.

To make a more accurate estimate of the confidence intervals in the tails of the distribution for the latency analysis, a pointwise CDF bound can be established. Given that the empirical CDF at point $x$ represents the proportion of observations less than or equal to $x$ (A.1), this value can be modeled as a binomial proportion with unknown parameter $p = F(x)$ and number of trials equal to the sample size $n$. Thus, the confidence interval for a given point on the CDF, e.g., 0.9, 0.99, or 0.999, can be calculated using the Wilson score interval~\cite{Wilson} as

\begin{equation}
\varepsilon_{Wilson|x}=\frac{1}{1+z_\alpha^2 / n}\left(\hat{p}+\frac{z_\alpha^2}{2 n} \pm \frac{z_\alpha}{2 n} \sqrt{4 n \hat{p}(1-\hat{p})+z_\alpha^2}\right),
\end{equation}

where $z_\alpha$ is the standard normal interval half-width for the probability $1 - \alpha$, and $\hat{p}$ stands for $k/n$ with $k$ being observations less than or equal to $x$.

Based on the Wilson score interval, Table~VIII shows the confidence intervals in the latency CCDFs for reliabilities of 90\%, 99\%, and 99.9\%. First, it should be noted that the confidence intervals are narrower than in the case of the DKW inequality, which is to be expected given that this inequality established an upper bound for the complete CDF. Second, it should be noted that the confidence intervals for $x = 0.999$ (equivalent to $10^{-3}$ in the CCDF) are in the range of $10^{-4}$, which validates the statistical confidence given the narrow confidence interval in this region of the distribution. Finally, it should be noted that there is no confidence interval in the tail of operator A because the out-of-coverage probability was above 1\%.

\renewcommand{\arraystretch}{1.5}
\setlength{\tabcolsep}{4pt}

\begin{table}[!t]
\centering
\caption{Wilson score interval - Confidence intervals for latency CCDF with $\alpha = 0.01$}
\label{tab:confidence_interval_wilson}
\resizebox{1\columnwidth}{!}{%
\begin{tabular}{|c||c||c||c||c||c|}
\hline\hline
\begin{tabular}[c]{@{}c@{}}Type of Connectivity\end{tabular} &
  \begin{tabular}[c]{@{}c@{}}\thead{SC - Operator A \\ 4G/5G}\end{tabular} &
  \begin{tabular}[c]{@{}c@{}}\thead{SC - Operator B \\ 4G/5G}\end{tabular} &
  \begin{tabular}[c]{@{}c@{}}\thead{SC - Satellite \\ Starlink}\end{tabular} &
  \begin{tabular}[c]{@{}c@{}}\thead{MC - Cellular \\ 4G/5G and 4G/5G}\end{tabular} &
  \begin{tabular}[c]{@{}c@{}}\thead{MC - Cellular Satellite \\ 4G/5G and Starlink}\end{tabular} \\ \hline\hline
Confidence Interval ($\varepsilon_{Wilson|x = 0.9}$) &
  $3.3 \times 10^{-3}$ &
  $3.1 \times 10^{-3}$ &
  $3.2 \times 10^{-3}$ &
  $2.9 \times 10^{-3}$ &
  $3.4 \times 10^{-3}$ \\ \hline
Confidence Interval ($\varepsilon_{Wilson|x = 0.99}$) &
  $\sim$ &
  $1.1 \times 10^{-3}$ &
  $1.1 \times 10^{-3}$ &
  $1.0 \times 10^{-3}$ &
  $1.2 \times 10^{-3}$ \\ \hline
Confidence Interval ($\varepsilon_{Wilson|x = 0.999}$) &
  $\sim$ &
  $3.8 \times 10^{-4}$ &
  $4.0 \times 10^{-4}$ &
  $3.6 \times 10^{-4}$ &
  $3.9 \times 10^{-4}$ \\ \hline\hline
\end{tabular}%
}
\end{table}

The results in this Appendix A statistically validate the analysis presented throughout the paper, which is possible due to the extensive measurement campaign carried out in rural environments where tens of thousands of ICMP (\textit{ping}) packets and tens of millions of UDP (\textit{iperf3}) packets have been acquired for evaluation of each technology in single-connectivity and multi-connectivity modes.

\bibliographystyle{IEEEtran}

\end{document}